\documentstyle[11pt,aaspp4]{article}

\def\wig#1{\mathrel{\hbox{\hbox to 0pt{%
          \lower.5ex\hbox{$\sim$}\hss}\raise.4ex\hbox{$#1$}}}}

\def\etal{{\it et~al.\,}}

\def\Dwa{$\,$\uppercase\expandafter{\romannumeral5}$\,$}

\def\sles{\lower2pt\hbox{$\buildrel {\scriptstyle <}
   \over {\scriptstyle\sim}$}}
\def\sgreat{\lower2pt\hbox{$\buildrel {\scriptstyle >}
   \over {\scriptstyle\sim}$}}

\newcommand{\Ch}{Chandrasekhar~}

\newcommand{\subCh}{sub-Chandrasekhar~}

\newcommand{\kh}{Kelvin-Helmholtz~}
\newcommand{\rt}{Rayleigh-Taylor~}

\newcommand{\Lsun}{$L_{\odot}$~}
\newcommand{\Lsunn}{$L_{\odot}$}
\newcommand{\Msun}{$M_{\odot}$~}
\newcommand{\Msunn}{$M_{\odot}$}
\newcommand{\Rsun}{$R_{\odot}$~}
\newcommand{\Rsunn}{$R_{\odot}$}

\newcommand{\kmsec}{km s$^{-1}$~}
\newcommand{\kmsecc}{km s$^{-1}$}

\newcommand{\Msunperyear}{$M_{\odot}$ yr$^{-1}$}
\newcommand{\dynescmcm}{dynes cm$^{-2}$}

\newcommand{\SNIa}{SNIa~}

\lefthead{Marietta \etal}
\righthead{Supernovae}
\slugcomment{Submitted to the Ap. J.}
\slugcomment{University of Arizona Theoretical Astrophysics Program preprint 00-00}
\slugcomment{\today}

\begin{document}

\title{Type Ia Supernova Explosions in Binary Systems: \\
The Impact on the Secondary Star and its Consequences}

\author{E. Marietta\altaffilmark{1}, Adam Burrows\altaffilmark{1}, and
        Bruce Fryxell \altaffilmark{2} }

\altaffiltext{1}{Department of Astronomy and Steward Observatory, 
                 The University of Arizona, Tucson, AZ \ 85721}

\altaffiltext{2}{Enrico Fermi Institute, 
                 The University of Chicago, Chicago, IL \ 60637}

\begin{abstract}

One method of discriminating between the many Type Ia 
progenitor scenarios is by searching for contaminating hydrogen and helium
stripped from the companion star. 
However, this requires
understanding the effect of the impact of the supernova shell on different
companion stars to predict the amount of mass
stripped and its distribution in velocity and
solid angle for the types of binary scenarios that
have been proposed as Type Ia progenitor models.

We present several high-resolution 2-D numerical simulations
of the impact of a Type Ia supernova explosion with hydrogen-rich main sequence, 
subgiant, and red giant companions. The binary parameters were chosen
to represent several classes of single-degenerate Type Ia progenitor 
models that have been suggested in the literature. 
We use realistic stellar models and supernova debris profiles 
to represent each binary system.
For each scenario, we explore the hydrodynamics of the
supernova-secondary interaction, calculate the amount of stellar
material stripped from the secondary and the kick delivered by
the impact, and construct the velocity and solid angle distributions of the 
stripped material. 

We find that the main sequence and subgiant companions lose $0.15-0.17$ \Msun
as a result of the impact of the supernova shell, $15\%$ of their mass.
The red giant companions lose $0.53-0.54$ \Msunn, $96\%-98\%$ of their envelopes.
The characteristic velocity of
the stripped hydrogen is less than $10^{3}$ \kmsec for all the scenarios:
$420 - 590$ \kmsec for the red giant companions,
$820$ \kmsec for the main sequence companion, and $890$ \kmsec 
for the subgiant companion.
The stripped hydrogen and helium contaminate a wide solid angle behind the
companion: $115^{\circ}$ from the downstream axis for the red giant,
$66^{\circ}$ for the main sequence star, and $72^{\circ}$ for
the subgiant.
With such low velocities, the bulk of the stripped hydrogen and helium
is embedded within the low-velocity iron of the supernova ejecta.
The hydrogen and helium may be visible in the late-time spectra as narrow 
emission lines. 

Although most of the stripped material is ejected at low velocities, all
the numerical simulations yield a small high-velocity tail.
The main sequence, subgiant, and the red giant companions are just 
under the Della Valle \etal (1996) upper limit from observations of SN 1990M 
taken near maximum light.

The main sequence companion receives a kick of $86$ \kmsec and the
subgiant receives a kick of $49$ \kmsecc.  In all cases, the kick to
the remnant is smaller than the original orbital velocity.
Because it is too small to intercept more than a negligible amount of momentum,
the red giant core will not receive an appreciable kick.

The impact of the supernova ejecta with the secondary star creates
a hole in the supernova debris 
with an angular size of  $\sim 30^{\circ}$ in the high-velocity ejecta and 
with an angular size of $\sim 40^{\circ}$ in the low-velocity ejecta.
This corresponds to $7-12\%$ of the ejecta's surface.
Because we explore binary scenarios that are close enough, or almost close enough, 
to be in Roche lobe overflow, the degree of asymmetry is similar for all the models.
The asymmetry in the supernova debris could have observational
consequences beyond the change in morphology of the supernova remnant. 
The asymmetry in the supernova atmosphere will result in distorted P Cygni 
profiles that might indicate the presence
of the companion star, but it will be difficult to use the degree of
asymmetry alone to discriminate between a main sequence, subgiant, or
red giant companion.

The impact of the supernova shell will have consequences for the
future evolution of the secondary star. After the impact, the main sequence star
is puffed up, much like a pre-main sequence star. The luminosity will rise
dramatically, to as high as $\sim5000$ \Lsunn,
as the extended envelope relaxes back into thermal equilibrium.
The subgiant companion will follow a 
similar sequence of events. The star will not be contaminated by much
supernova debris from the initial impact, but it may accrete
low-velocity iron-group elements (or oxygen and silicon if the
ejecta is radially mixed) at late times.

A He pre-white dwarf will be left behind after a supernova explosion
with a red giant companion. Almost all the envelope will
be ejected by the impact, but a residual amount of material ($\sim 0.02$ \Msunn)
will form an extended, hydrogen-rich envelope around the degenerate core.
The star will evolve away from the red giant branch 
on a timescale of $10^{5}-10^{6}$ years. On its post-RG track,
it may appear as an underluminous O or B star before passing
through an sdO or sdB phase on its way to a 
standard He white dwarf cooling track.
This binary scenario could be a possible pathway for the formation
of a subset of single, low-mass He white dwarfs.

\end{abstract}

\keywords{supernovae: general}

\newpage

\section{Introduction}

Unlike a Type II explosion, a Type Ia explosion, the thermonuclear 
explosion of a CO white dwarf, is characterized by the absence of
hydrogen in its spectrum.
Although a massive star that is responsible for a 
Type II supernova explosion can be in a single or a binary system, 
the Type Ia progenitor scenarios that are based on the
thermonuclear explosion of an accreting white dwarf require a companion 
as a mass donor to enable the white dwarf to reach its critical mass.
However, despite searches for the remnant of the companion star, 
no direct evidence has yet been observed (\cite{ruiz97}).

One method of discriminating between the many Type Ia 
progenitor scenarios is by searching for contaminating hydrogen and helium
stripped from the companion star. However, this requires
understanding the effect of the impact of the supernova blast on different
companion stars to predict the amount of hydrogen
stripped and its distribution in velocity and
solid angle for the types of binary scenarios that
have been proposed as progenitor models.
To date, there have been only three numerical investigations
of supernova impacts on companion stars.
Fryxell \& Arnett (1981) and Taam \& Fryxell (1984) focused on impacts with 
low-mass main sequence stars.  
Livne \etal (1992) focused on red giant companions.
In this paper, we improve upon the earlier theoretical work
by exploring the nature of the interaction between the supernova ejecta 
and the companion star for specific Type Ia binary scenarios
with higher resolution than was possible in the past.
We select realistic ejecta profiles and realistic stellar models. 
It is no longer necessary to represent the
supernova-secondary interaction  by the collision of a simple shell with a
polytropic companion.
Although the earlier simulations studied the stripped mass and kick,
they did not study or provide the
distribution of stripped hydrogen in velocity and solid angle,
which are essential to predict any spectral signatures of the impact.

The goal of this project is to simulate with a 2-D hydrodynamics code
the impact of a realistic supernova blast on a companion star in the 
context of current Type Ia progenitor scenarios in which the companion
is hydrogen-rich.
We select several of these scenarios from the literature 
to cover the possible range of secondary types which are
typically low-mass ($1.0-2.0$ \Msunn) main sequence, subgiant, 
or red giant stars. We use realistic stellar models and supernova
debris profiles to represent each binary system.
For each scenario we explore the hydrodynamics of the
supernova-secondary interaction, calculate the amount of stellar
material stripped from the secondary and the kick delivered by
the impact, and construct the velocity and solid angle distributions of the 
stripped material. With the velocity and spatial distributions of the 
stripped material within the supernova debris
it is possible to predict whether the contaminating hydrogen and helium
are observable in the supernova spectrum as 
broad emission lines near maximum light
or as narrow emission lines at late times.
In addition, we can explore the change in structure of the
secondary star caused by the impact to explore the possible
final states (perhaps observable) of the secondary.

We find that our main sequence and subgiant companions lose $0.15-0.17$ \Msun
as a result of the impact of the supernova ejecta.
In contrast, the red giant companions lose almost their entire envelopes,
$0.53-0.54$ \Msunn.
The bulk of the stripped hydrogen, which has a characteristic
velocity of less than or equal to $10^{3}$ \kmsecc, is embedded within
the inner, low-velocity supernova debris.
The complication that hydrogen and helium can be hidden deep and
unseen within the ejecta, and obscured by competing
iron and cobalt lines, can be addressed by radiative
transfer calculations (\cite{pinto99}).
In principle, radiative transfer calculations can
reveal how much of the stripped material can be hidden
and when it is likely to be observable.

\subsection{History of Theory}

The effect of a supernova explosion on a nearby companion star
has been the subject of speculation for many years. 
One issue raised was the strength of the kick 
delivered to the companion. 
Colgate (1970) proposed that the shock-heating
of the stellar envelope would cause material to evaporate
preferentially back in the direction of the supernova.
The secondary would receive a substantial kick from the
ablation (evaporation from the surface of the star), 
as well as a smaller kick from the direct collision with the ejecta. 
Other groups (\cite{cheng74}; \cite{wheeler75}) made similar arguments
supporting the suggestion that the kick due to ablation was
larger than the kick from the collision.
Numerical simulations by Fryxell \& Arnett (1981) and Taam \& Fryxell (1984)
showed that the momentum transfer to the secondary
was much less than previously thought and that 
ablation did not substantially increase the kick.
The small binary separations we employ for the
simulations in this paper, necessary for
Roche lobe overflow, result in kicks
larger than the $\sim 10$ \kmsec kicks reported
by Fryxell \& Arnett (1981). However, our
typical $40-80$ \kmsec kicks are still small
in comparison with typical orbital velocities.

The impact of the supernova debris on a nearby companion
may be quite dramatic. The supernova ejecta may 
either directly strip material from the 
companion by direct transfer of momentum, or, 
through the conversion of the blast kinetic energy into
internal heat, by evaporation.
Because the incident energy (a fraction of $10^{51}$ ergs) greatly 
exceeds the secondary's binding energy ($10^{48}$ ergs),
Sofia (1967) suggested that a main sequence secondary 
in Roche lobe overflow would be completely destroyed.
It was not until a more detailed study of the shock propagation
into the stellar envelope was done that it was
realized that, because the kinetic energy of the impact is
deposited in the outer layers of the star, a 
main sequence star could easily survive such an impact (\cite{cheng74}).

Wheeler \etal (1975) analytically estimated the amount of 
mass stripped and the kick received by the companion
as a result of the inelastic collision and the shock-heating.
They calculated the outcome of the impact for an arbitrary secondary star
using a geometrical parameter they devised, which depended on
the incident momentum and the mass and compactness of
the secondary.
Their analysis implied that
main sequence stars (represented by an $n=3$ polytrope), 
being tightly bound, are only slightly 
stripped by an explosion in a close binary system.
Red giants with loosely-bound envelopes are so catastrophically 
heated by the impact that the entire envelope is ejected.

The analytic prescription for the stripped mass and kick
proposed by Wheeler \etal (1975) was put to the test 
in a series of numerical simulations of supernova impacts
on both low-mass main sequence companions
(\cite{fryxell81}; \cite{taam84}) and
low-mass red giants (\cite{livne92}).
Fryxell \& Arnett (1981) demonstrated that the collision of
a supernova shell with a $2.0$ \Msunn, n=$3$ polytrope,
with a binary separation $5.9$ times the stellar radius,
ejects $0.013 - 0.052$ \Msun from the secondary,
a mass loss roughly consistent with the
analytic estimates of Wheeler \etal (1975).
Although in their study the emphasis was on the momentum transfer
rather than the stripped mass, Taam \& Fryxell (1984)
performed numerical simulations with
n$=3/2$ and n$=3$ polytropes that were designed
to represent the range of compactness expected from a 
fully convective star to a radiative star.
They found that the momentum was more efficiently transferred for
the n=$3/2$ star, which is less centrally concentrated than the
n$=3$ star. The stripped n$=3/2$ star presents a
larger geometrical area to intercept the momentum, even
when a higher fraction of mass is stripped, so that the
momentum transfer is more efficient.
More recently, Livne \etal (1992) simulated a Type Ia supernova 
explosion of a white dwarf in a close binary 
with a low-mass red giant that is almost close enough to be in Roche lobe
overflow. The blast stripped the red giant of almost its entire 
envelope, imparting a velocity to the stripped material well below 
the velocity of the supernova ejecta. This not only confirmed 
the conclusion of Wheeler \etal (1975) that the impact would destroy
the red giant, but also verified a prediction of Chugai (1986).
Chugai (1986) focused on the supernova red giant interaction for
Type Ia events in which the hydrogen donor is a symbiotic star - 
with a binary separation $3-10$ times the red giant's radius.
Chugai's analytic model predicted that $0.3$-$1.0$
\Msun would be ejected from the companion with a characteristic
velocity less than or equal to $10^{3}$ \kmsecc, much smaller than the
$10^{4}$ \kmsec characteristic velocity of the supernova ejecta,
and that it would fill the inner $20 \%$ of the radius of the supernova ejecta.
The confirmation by Livne \etal (1992) of the low velocity 
of the stripped hydrogen has important implications for the identification
of Type Ia progenitors.
Our more detailed calculations build on and extend this earlier, seminal work.

\subsection{Discriminating Between Type Ia Progenitor Scenarios
            by Searching for Hydrogen}

The exact amount of hydrogen stripped and its characteristic 
velocity is a key issue in light of the current interest in 
identifying the progenitors of  Type Ia supernovae.
Because it is difficult to reconcile the extremely hydrogen-poor
spectra with the presence of a hydrogen-rich companion,
the amount of stripped hydrogen can be used
to discriminate between Type Ia progenitor scenarios.
Type Ia explosions, which are characterized by a
lack of hydrogen,
could be the result of mass accretion onto a white dwarf in a 
binary system (single-degenerate scenario) (\cite{whelan73}; \cite{nomoto82a}) 
or the merging of two degenerate white dwarf stars (double-degenerate scenario) 
(\cite{iben84}; \cite{webbink84}). 
Naturally, the double-degenerate scenario could account for the lack of hydrogen 
in the supernova spectrum. However, no double-degenerate binaries have
been found that are close enough to merge within a Hubble time and
are massive enough to exceed the \Ch mass 
(\cite{marsh95}; \cite{saffer98}).
The single-degenerate scenario
is currently in favor (\cite{livio99}).
If the companion is a main sequence, subgiant, or red giant star,
the single-degenerate scenario necessarily implies that the binary system 
is rich in hydrogen. Hydrogen-rich material could be ejected from the 
secondary star as a consequence of the impact of the supernova shell.
Hydrogen could also be present in the immediate environment as circumstellar 
material from stellar winds,  mass lost from the primary in an earlier 
phase of mass transfer, or even as a layer of hydrogen on the white dwarf primary. 
If the hydrogen is swept up by the supernova ejecta, the origin
of the material may determine its characteristic velocity, which
in turn determines when it is most likely to be observed.
Circumstellar material swept up with the supernova ejecta
is expected to be observed as transient narrow
H$_\alpha$ emission or absorption lines (\cite{wheeler92}; \cite{filippenko97})
near maximum light. If the hydrogen is stripped from the secondary and
embedded within the inner iron layer of the supernova ejecta, it is
more likely to be observed in emission as narrow H$_{\alpha}$ lines 
months after maximum light (\cite{chugai86}).
Searches for hydrogen must be targeted at either early-times near
maximum light or at late-times when the photosphere has receded to
reveal the iron layer (\cite{chugai86}).

Identification of hydrogen at early-times, 
either in emission or absorption,
has been claimed in at least three Type Ia supernovae: 
1981B, 1990M, and 1990N.
Branch \etal (1983) presented high-quality optical spectra of
SN1981B from maximum light to 116 days post-maximum.
Their March 13 observation showed a small, narrow emission feature
consistent with the rest wavelength of H$_{\alpha}$. The feature
was absent from a spectrum that was taken just 5 days later.
Unable to firmly conclude that the
feature was H$_{\alpha}$, they suggested that high-resolution spectra
be taken of supernovae near the same phase ($\sim 6$ days
past maximum light) to search for H$_{\alpha}$ emission.
Cumming \etal (1996) searched for narrow H$_{\alpha}$ emission in 
high-resolution spectra of SNIa 1994D 10 days before
maximum and 6.5 days after maximum without detecting any hydrogen
either in emission or absorption. 
Likewise, Ho \& Filippenko (1995) in their high-resolution echelle 
observations of SNIa 1994D at 23 days past maximum
did not detect any H$_{\alpha}$ in emission or absorption.

From a series of observations of SNIa 1990N from
14 days before maximum light to 1 week past maximum light,
Leibundgut \etal (1991) suggested that an unidentified absorption
feature at $6300$\AA\ could be hydrogen.
They suggested that a thin layer of hydrogen on the surface of 
the progenitor star could be responsible for the feature, but 
because the velocity of the feature ($1.2\times10^{4}$ \kmsecc) was
much lower than that of the other lines, this was unlikely.
Later, based on an analysis of optical and UV spectra of SN 1990N 
using synthetic spectra,
Jeffery \etal (1992) concluded that the unidentified
absorption feature at $6300$\AA, and an additional unidentified feature
at $6900$\AA, were probably due to \ion{C}{2} at $6580$\AA\  
and $7243$\AA\ .

Polcaro \& Viotti (1991) identified a broad H$_{\alpha}$ absorption
feature in SNIa 1990M at four days after maximum light. This
was later convincingly ruled out by Della Valle \etal (1996), who showed
from a spectrum obtained a few days past maximum light
that the absorption feature was due solely to observational bias.
They found no evidence of H$_{\alpha}$ in SN 1990M. 
Using an upper limit to the H$_{\alpha}$ equivalent width and a 
measurement of the \ion{Si}{2} $6350$\AA\  equivalent width, 
they set an upper limit in the supernova atmosphere 
for H/Si of $2.0 \times 10^{-6}$,
relative to solar.
By assuming perfect mixing and using the abundances of the standard
W7 model (\cite{nomoto84}), they derived
the upper limit to the mass fraction, X(H), of
$0.007$. However, they pointed out that if the hydrogen
is  buried deep within the supernova ejecta instead of being
perfectly mixed, the integral mass fraction could be much higher.
Likewise, if the hydrogen is restricted to the outer layers, 
the mass fraction could be much lower.
Scaling X(H) by the ratio of the mass in the
atmosphere to the total mass and assuming that 20 days past the explosion
only $1/20$ of the ejecta are revealed, they 
derived an X(H) of $\sim 0.007\ M_{atm}/M_{tot} \approx 3 \times 10^{-4}$.

A similar argument was made earlier by Applegate \& Terman (1989) to 
place an upper limit on the hydrogen mass fraction in SN 1981B.
Using the upper limit of $ \left[  {\rm H/Si} \right] < -4.0$ in the
LTE spectral fit to the maximum light observations of
SN 1981B of Branch \etal (1982), Applegate \& Terman (1989) derived an
upper limit on the hydrogen mass fraction near maximum light
in SN 1981B of $0.02$.
To find a lower limit to the hydrogen contamination
due to the impact of the
supernova ejecta on the secondary star, Applegate \& Terman 
used the analytic prescription developed by Wheeler \etal (1975).
They focused on a cataclysmic variable system with a
$0.2$ \Msun main sequence companion,
represented by an n=$3/2$ polytrope, that
was close enough to the primary to be in Roche lobe
overflow. If the Type Ia progenitors are cataclysmic variables,
\cite{applegate89} concluded that X(H) $> 0.01$.
This is just below the Applegate \& Terman upper limit of X(H) $< 0.02$.
In contrast, if the companion is a low-mass red giant with an
envelope mass of $0.5$ \Msunn, then X(H) $> 0.20$, which clearly
would exceed the upper limit they deduced from observations of SN 1981B.
Applegate \& Terman (1989) conclude that binary scenarios with hydrogen-rich 
companions are not likely candidates for Type Ia progenitors.
However, to directly compare their lower limit
with their upper limit based on maximum light observations of SN 1981B, 
they implicitly assumed that the stripped hydrogen from
either the low-mass main sequence star or the red giant
companion was ejected at high velocities.

It is currently believed, based on the work of Chugai (1986)
and Livne \etal (1992), that any material
stripped from a red giant companion will have a characteristic 
velocity $\lesssim 10^3$ \kmsecc, much lower than that of the 
supernova ejecta. 
As we show below, we agree with this general conclusion.
If any hydrogen is embedded within the supernova
ejecta, narrow hydrogen emission lines will appear only after the 
photosphere retreats into the iron layer.
Thus, a critical test of the presence of a secondary star
is the detection of low-velocity hydrogen lines in the late-time
spectrum (\cite{ruiz93}). However, such a detection is difficult 
because of the number of Fe lines that overwhelm the spectrum.
To date, there has been only one claim of a detection of a
low-velocity hydrogen line, ostensibly from a stripped companion.
Ruiz-Lapuente \etal (1993) initially identified a weak H$_{\alpha}$ line
in a spectrum of SNIa 1991bg, a peculiar underluminous Type Ia supernova, 
197 days after maximum light.
However, this was not verified by Turatto \etal (1996). 
Turatto \etal were not able to associate any emission lines
with H$_{\alpha}$ in their observations of SN 1991bg,
although they suggested a possible blend with 
[\ion{Co}{3}] at $6578$\AA\  or
numerous \ion{Fe}{2} and [\ion{Fe}{2}] lines in that region of the spectrum.
Garnavich \& Challis (1997), in a reanalysis of a spectrum of SN 1991bg at
200 days after maximum light, confirmed the existence of the the narrow
emission lines of Ruiz-Lapuente \etal (1993), but they could not unambiguously
identify them.

A key question for theoreticians is how much contaminating
hydrogen and helium, whether of circumstellar or companion origin,
can be hidden within the supernova ejecta.
Very few theoretical upper limits have been set 
(\cite{wheeler90}; \cite{wheeler92}).
Wheeler (1992) claimed that, if the hydrogen is in LTE
in the outer layers of the supernova ejecta, as much as
$0.1$ \Msun could be present without contributing to the
spectrum at maximum light. 
Branch \etal (1991) found a tighter upper limit of $1 \%$ by mass 
($\sim 0.01$ \Msunn) in a non-LTE calculation. 
By adding hydrogen to a synthetic spectrum and comparing with the
maximum-light optical spectrum of SN 1981B, 
they concluded that $1\%$ by mass is an upper limit on the 
hydrogen contamination for a layer of hydrogen uniformly mixed 
in the line-forming layer at maximum light. 
They found that, not only would a weak H$_{\alpha}$ line be visible, but,
as an indirect consequence of hydrogen contamination, that
the optical depth in the Lyman continuum would change the ultraviolet
radiation field, the ionization structure in the line-forming layer, 
and, hence, the strengths of many of the non-hydrogen lines.
Because Applegate \& Terman (1989) estimated X(H) $> 0.01$,
Branch \etal (1991) concluded that accreting white dwarfs with hydrogen-rich 
companions are unlikely to be progenitors of Type Ia supernovae.
However, there are several caveats.
First, the synthetic spectra employed for these
investigations were done only near maximum light, 
when only high-velocity hydrogen would be visible. 
If most of the contaminating
hydrogen is at low velocities, synthetic
spectra need to be calculated at
much later times when the photosphere
has receded into the inner ejecta. Second, the hydrogen buried within
the supernova ejecta may not be uniformly mixed in angle. 
Because non-uniform mixing may act to hide the embedded hydrogen,
the angular dependence of the spectra needs to be considered. 

In this paper, we explore in detail the theoretical expectations
for the distribution of hydrogen and helium in Type Ia debris, the 
hydrodynamic character of the impact, the kick to the
secondary, and the nature of the post-impact structure.
In \S \ref{sec:numericalmethods}, we review the 
numerical methods used. In \S \ref{sec:biscenario},
we describe the classes of supernova 
Type Ia progenitor scenarios and the criteria for selecting
the candidate scenarios.
In \S \ref{sec:mainsequence}, we discuss details of the main sequence 
simulations, including the hydrodynamics of the impact,
the stripped mass, the velocity and solid angle distributions of the 
stripped material, and the kick received by the companion.
In \S \ref{sec:separation}, we present a sequence of simulations
with the main sequence secondary to determine systematic trends 
in the quantity of stripped hydrogen and its velocity and
solid angle distributions as the binary separation increases.
In \S \ref{sec:subgiant}, we discuss details of the subgiant
simulations, including a description of the impact,
the stripped mass, the velocity and solid angle distributions of the 
stripped material, and the kick received by the companion.
Section \ref{sec:redgiant} contains similar details 
for the red giant simulations.
In \S \ref{sec:spectrum}, we explore the implications of the
revealed velocity distributions of the stripped material and the
possibility that contaminating hydrogen may be observable
in a Type Ia supernova spectrum.
In \S \ref{sec:postimpact}, we speculate on the effect of the
impact on the future evolution of the secondary, and
in \S \ref{sec:conclusions} we summarize our
conclusions and suggest directions for future work.
Postscript images, MPEG movies, and a selection of figures
presented in this paper, are posted at http://www.astrophysics.arizona.edu and are
available via FTP at www.astrophysics.arizona.edu, in directory pub/marietta.

\section{Numerical Methods and Techniques}
\label{sec:numericalmethods}

The hydrodynamics code we employ for these simulations
is an extension of the code, {\it Prometheus} (\cite{fryxell89}),
which is based on the Piecewise-Parabolic Method (PPM) 
(\cite{colella84}; \cite{colella85}).
Our version of PPM is 
a non-relativistic, explicit, automatically conservative, Eulerian
scheme that achieves second-order spatial and temporal accuracy. 
Fluxes at interfaces are obtained by solving the Riemann 
shock-tube problem and shocks are resolved to one or two zones. 

A simple equation of state, including just radiation and ideal gas, is
employed for all the calculations involving the main sequence
companion and the envelope of the red giant; the degenerate core of
the red giant is added only as a gravitational point mass.
For the calculations involving subgiant companions,
where degenerate electrons are required,
we use a tabulated equation of state with arbitrary degeneracy and relativity.
We tabulate the pressure, energy, and entropy of the ionization
electrons, and any pair-produced electrons and positrons, by directly
integrating the appropriate Fermi-Dirac integrals,
following the prescription of Cox \& Giuli (1968).
Using the ionization electron density
($n_{e} = \rho Y_{e} N_{av}$, where $\rho$ is the density, $Y_{e}$ is
the electron fraction, and $N_{av}$ is Avogadro's number)
and the temperature ($T$) as independent variables,
we use an iterative technique to solve for the degeneracy factor
($\eta = \mu/ k_{B}T$, where $\mu$ is the chemical potential, and
$k_{B}$ is Boltzmann's constant),
which can be used to directly integrate
the Fermi-Dirac integrals for pressure and energy.
This method, although laborious, is
thermodynamically consistent. The table, once generated,
is accessed by a second-order interpolation routine. We estimate
the pressures and energies to be accurate to better than $\sim 10^{-4}$.

We incorporate into PPM an integral Poisson solver created by M\"uller \& Steinmetz (1995)
to calculate the gravitational potential for an arbitrary 2-D mass 
distribution. Having found the gravitational potential, the gravitational 
force can be calculated using a finite difference approximation to the gradient.
In the simulations involving the red giant companions, the 
gravitational point mass of the compact core is added directly to the force.
In our implementation the solver expands the gravitational potential
in Legendre polynomials. Up to twenty moments can be employed, including
the dipole moment which we include to allow the secondary to move freely 
down the hydrodynamic grid in response to the impact of the blast.
With the addition of an interpolation routine, the hydrodynamical 
equations can be differenced in spherical $(r,\theta)$, Cartesian $(x,y)$,
or cylindrical coordinates $(\rho,z)$ 
while the gravitational potential can be constructed in spherical coordinates $(r,\theta)$.
To improve accuracy, the gravitational potential can be centered
on the center of mass at each cycle, in effect following the star 
as it moves down the grid.
This decreases errors, notably in the stellar core, that over the 
long timescale of the simulation can otherwise slowly decrease the star's
terminal velocity. Even with this improvement,
the terminal velocity slowly decreases at an average rate of $\sim 2.8\%$ per stellar
sound crossing time in a typical simulation with a low-mass main sequence companion. 
Thus, as will be seen in \S \ref{sec:kickms}, we determine the 
kick after the terminal velocity is reached ($4000-5000$ seconds), but 
before the small errors have time to grow appreciably.

For simplicity, we use all the mass on the hydrodynamic grid, stellar
and supernova, to calculate the gravitational potential.
Because the grid is so large, typically $6$R (R = the stellar radius) 
in the $\rho$ direction,
and, since the explosion is $\sim 1$R away from the upper boundary,
about $42\%$ of the supernova debris passes through the grid.
Only a small
fraction of the ejecta directly collides with the secondary. 
The characteristic velocity of the supernova ejecta is so fast
that the error to the kick from including the
supernova material on the grid is only  $\sim 4$ \kmsecc, which is
an error of $\sim 5 \%$.

For this project, a selection of low-mass stellar models were evolved by
Chaboyer (1998).
Following the lead of Sills \etal (1997), we reintegrate the 
main sequence and subgiant secondaries using a
1-D fourth-order Runge-Kutta program, the entropy and composition
profiles, and the same equation of state used in our PPM code.
We do not attempt to reproduce the distortions in the secondary 
due to the Roche lobe geometry. 

The weakly-bound envelope of the red giant requires special treatment.
We reintegrate its entropy profile
with a softened potential ($\phi(r) = -G M_c / ( r + r_{c} )$)
to help stabilize it.
Even with this adaptation, we find it necessary to
switch to spherical coordinates for the red giant
impacts. We surround each secondary star with
a high-entropy, very low-density hydrostatic envelope which serves to
fill the Eulerian grid with a background medium.
In later sections we refer to this background material as the 
``circumstellar'' medium to distinguish it from the stellar envelope.
The new 1-D models with their hydrostatic envelopes are interpolated 
onto a 2-D cylindrical grid (main sequence or subgiant models)
or a 2-D spherical grid (red giant envelopes).
We use 2-D cylindrical coordinates whenever possible in order to
efficiently follow the stripped material.
To stabilize the envelope of the red giant, we must use
2-D spherical coordinates. 

We verify that each secondary star remains in hydrostatic
equilibrium by running a simulation without the supernova explosion. 
The main sequence and subgiant simulations were run for three
sound crossing times. Because the envelope of the red giant will be
completely disrupted in less than one sound crossing time, the red giant 
simulation was run for only one sound crossing time. 
We estimate the fractional change in radius 
by the fractional change in gravitational energy
for the main sequence and subgiant secondaries.
We find that the radius changes by at most $\sim1\%$.
In the red giant case, the radius changes by at most $\sim 8\%$.

We perform all of the simulations with the main sequence
and subgiant secondaries in 2-D using cylindrical
coordinates ($\rho,z$), with the z-axis defined to be
the direction joining the primary and secondary of the
original binary system.
We estimate that the error in neglecting the orbital motion of the secondary,
necessary for a 2-D calculation, is $\sim10^{-2}$, the ratio of a typical
orbital velocity of $10^{2}$ \kmsec to a characteristic
velocity of the supernova ejecta of $10^{4}$ \kmsecc.
The thermonuclear explosion of the white dwarf occurs exterior
to the grid and the supernova ejecta, specified by a realistic 
density and velocity profile, flows onto the grid via
time-dependent boundary conditions in assumed spherical
homologous flow. The other exterior boundaries are subject to the
constraint that material is allowed only to leave the grid. 
The interior boundary ($\rho = 0$) is reflecting.

We switch to spherical coordinates for the red giant companions
because only in these coordinates is the loosely-bound
envelope of the red giant stable.
We position the red giant envelope at the origin and 
add a gravitational point mass to the force to represent the 
degenerate core.
We assume throughout the simulation that the
degenerate core does not move as a result of the impact. 
We justify this by noting that, because of the extremely small 
solid angle subtended by the degenerate core and its
very high areal density,
it can receive only a negligible ($\ll 1$ \kmsecc) kick.
The supernova explosion, which in this case occurs on the grid,
is interpolated onto the grid from a post-explosion ejecta profile. 
To avoid numerical problems associated with high-Mach flows, 
we assume an initial temperature in the supernova ejecta high enough 
that the internal energy is $\sim 7.7 \%$ of the specific kinetic energy.\footnote{
Our initial temperature is a factor of $1.0-4.0$ higher than the temperature
provided in the supernova ejecta model hedtb11 (\cite{woosley94}).}
To minimize numerical problems at the center of the supernova
where the expansion leaves a low-density interior,
we employ a minimum temperature which decreases
linearly with time to mimic the cooling of the interior.
Outside of the supernova ejecta at the interface between
the supernova ejecta and the ``circumstellar'' medium,
a minimum temperature of $100$ K is enforced.
These numerical problems appear only in the spherical
calculations. The cylindrical calculations in which 
the supernova is added by time-dependent boundary conditions
are always well-behaved.

With the change to spherical coordinates, we alter the boundary conditions.
The outer boundary is straightforward.
As in cylindrical coordinates, the outer radial boundary is
transmitting, but subject to the constraint that material is
allowed only to leave the grid. 
The inner radial boundary is
fixed at a small, but non-zero, radius. This helps to stabilize
the interior of the envelope and, in addition, helps to avoid 
Courant problems.
The inner boundary is non-transmitting
(no mass flux allowed) with a zero-velocity condition.
The inner boundary seems to affect the flow only after
the envelope has been stripped. 
After the mass-stripping phase of the interaction,
we regrid the calculation by removing the innermost zones,
usually the first $25$, in effect 
moving the inner boundary outward about a factor of 10 in
radius, which increases the timestep by the same factor,
allowing us to continue the simulations efficiently.
The mass enclosed in the innermost zones is added to the 
gravitational point mass. The region removed
is typically only $1 \%$ in radius, and therefore, only $10^{-4} \%$ in 
computational volume.
The boundary along the axis of symmetry is reflecting.
For high-velocity flow, the axis of symmetry 
manifests minor numerical artifacts which, however, do not
affect the conclusions of the work.
 
In addition to determining the stripped mass and the kick
given to the companion, we find the distribution
of the stripped material in both velocity and
solid angle. 
The velocity distribution can in principle be used
with a radiative transfer code to predict when, and if,
the contaminating hydrogen is observable in the supernova Type Ia 
spectrum and to place useful constraints on the type of binary scenarios
likely to be responsible for Type Ia supernovae.
The solid angle distribution can be used to determine by how much
the stellar material lags behind the supernova ejecta and
the size of the solid angle contaminated by the stellar hydrogen
and helium. It may also be relevant for estimates of the polarization
of the emergent light.
As stripped stellar material flows through the outer boundary, we
record its mass, velocity, entropy, and composition.
We advect the electron fraction Y$_{e}$ and ion fraction Y$_{i}$
along with ten general composition labels, five of which are reserved for
the Type I supernova ejecta 
(``hydrogen'', ``helium'', ``oxygen-group'', ``silicon'', and
``iron-group''), four for the companion star 
(``hydrogen'', ``helium'', ``oxygen-group'', and ``silicon/iron-group''), 
and one for the ``circumstellar'' medium. 
The Y$_{i}$ and Y$_{e}$ are required by the equation of state, but
we advect the composition labels as mass fractions so we can keep 
track of the origin of the stripped companion material and can distinguish it
from the supernova and ``circumstellar'' material with which it is intermixed.
To find the velocity and solid angle distributions,
we include the distribution of the stripped companion material on the grid,
as well as the stripped material that has left the grid. 
The stripped mass left on the grid is usually extremely small,
given the length of a typical simulation.

For the main sequence and subgiant simulations we generally employ a
$\sim 300 (\rho) \times 600 (z)$ cylindrical grid, and for the
red giant a $\sim 600 (r) \times 300 (\theta)$ spherical grid, 
with nonuniform zoning that is finest in the region surrounding the secondary
and coarsest near the edge of the grid.
To cover $6$ stellar radii around the secondary and still
resolve the center of the star well enough to
maintain hydrostatic equilibrium, nonuniform zoning is necessary. 
This nonuniform zoning, although efficient, creates
difficulties in quantifying the effect of increasing or
decreasing the resolution.

To test the effect of the resolution
we employ a cylindrical grid that has uniform zoning in a
rectangular region enclosing the secondary. Outside this region,
the zones are slowly increased in size to keep the
calculation tractable. 
For the main sequence calculation,
we performed two simulations: one with a spacing
of $8.0\times10^{3}$ km requiring a
$300 (\rho) \times 550 (z)$ grid
and one with a slightly
smaller spacing of $6.0\times10^{3}$ km
requiring a $320 (\rho) \times 845 (z)$ grid.
We can not vary the ``resolution'' by a large
factor because greatly decreasing the resolution
destabilizes the secondary while greatly increasing
the resolution creates enormous Courant and
data storage problems.

Nevertheless, we find that changing the resolution
in the region around the secondary can affect the results 
in several systematic ways.
First, the amount of mass stripped increases, but only very slightly,
as the resolution increases. 
In the high-resolution simulation the secondary lost
almost $\sim 0.1\%$ more mass than in the low-resolution
simulation. 
Second, as the resolution increases, the kick to the
remainder of the secondary increases.
The velocity of the center of mass follows the
same overall profile, the only difference being
a slight increase of $1.1\%$ in the terminal velocity.
An increase in momentum transfer with higher resolution
was noted by Taam \& Fryxell (1984). They attributed
this trend to a more accurate description of the momentum
transfer in the outer layers of the star, where
the shock energy is deposited, because of the steep density 
gradient there.
Third, as the resolution increases, the velocity 
of the stripped material systematically shifts
to slightly lower velocities. 
The velocity at the half-mass point 
shifts from $841$ \kmsec to $829$ \kmsecc.
If with higher resolution the
transfer of momentum to the remainder of the secondary
is more efficient, then less momentum 
is available for the stripped material. 
Apart from the subtle systematic shift as the resolution changes,
the features in the moderate-velocity
region ($1000 - 3000$ \kmsecc), which we associate
in \S \ref{sec:hydroms} with the composition transitions
in the supernova ejecta, vary in position
and amplitude.
However, the velocity distributions of the low- and 
high-resolution calculations show the same overall profile,
especially at the low- and high-velocity tails.
Above $3\times10^{3}$ \kmsecc, the low-resolution
distribution has $12\%$ more mass than the high-resolution
simulation. 
To reprise, as the resolution increases, the amount
of stripped mass and the kick imparted to the
secondary increase slightly, and the half-mass
velocity of the stripped material decreases slightly.

\section{Standard Type Ia Supernova Binary Scenarios}
\label{sec:biscenario}

A Type Ia supernova could be the product of a merger of
CO white dwarf with a He or CO white dwarf (double-degenerate scenario),
or the explosion of a white dwarf which has reached
its \Ch mass by accreting hydrogen or
helium from a nondegenerate companion (single-degenerate
scenario) (\cite{iben84}). 
These two competing binary scenarios originate from
very different evolutionary paths and naturally have
different predictions for the Type Ia supernova rates.
Binary population synthesis, a statistical tool
for exploring the possible phases of binary evolution,
can be used to estimate rates for progenitor scenarios
to discriminate between progenitor models
(cf.~\cite{iben84}; \cite{tutukov92}; \cite{branch95}).
However, the Type Ia supernova rates from population
synthesis calculations suffer from serious uncertainties.
Common envelope evolution is not well-understood.
It is parameterized by $\alpha_{CE}$, the
efficiency with which the orbital energy is used to
eject the common envelope, whose value is uncertain.
The realization frequencies are dependent on the
stellar population and age. Another complication 
arises from the different models for the
thermonuclear explosion of the white dwarf.
In a \Ch explosion (the standard model), the CO white dwarf 
accretes material until it approaches the \Ch mass and
carbon ignites in the core (\cite{woosley86}).
In \subCh explosions, the ignition occurs in a helium
layer around the CO white dwarf before the \Ch mass
is reached (\cite{nomoto82b}). 
In general, binary population synthesis calculations
favor the double-degenerate scenario over the single-degenerate scenario (\cite{branch95}).
Recent calculations by Branch \etal (1995) for a young population ($10^{8}$ yr)
find double-degenerate rates of $\sim 10^{-3}$ yr$^{-1}$.
This is closer to the current galactic Type Ia supernova
rate of $4.0 \times 10^{-3}$ yr$^{-1}$ than the
rates of $10^{-4} - 10^{-6}$ yr$^{-1}$ for each of the
many possible single-degenerate scenarios.
But, for an older population ($10^{10}$ yr)
the double-degenerate rate drops to
$\sim 10^{-4}$ yr$^{-1}$ which can be matched by
the realization frequency for symbiotic systems,
a single-degenerate scenario in which the white
dwarf accretes hydrogen from the wind of its
red giant companion.
Likewise, Ruiz-Lapuente (1996) finds that for an Sb galaxy
the double-degenerate rate is $\sim 10^{-4}$ yr$^{-1}$, 
but that several single-degenerate scenarios have rates
similar to this.
Because of the complexity of the population synthesis calculations, 
it is difficult to exclude any Type Ia progenitor scenario outright
just on the basis of realization frequencies.
The presence of hydrogen in Type Ia supernova spectra is a stronger 
indication of a single-degenerate progenitor system
because any Type Ia explosion in a double-degenerate binary system 
is expected to be hydrogen-free.
In this paper, we examine only the single-degenerate
scenarios for Type Ia supernova explosions.

Single-degenerate Type Ia models can be coarsely subdivided
(see Table \ref{singlemodels})
into Hydrogen Cataclysmic Variables (H CV), 
Hydrogen Cataclysmic-Like Variables (H CVL), 
Symbiotic Stars (SS), Hydrogen Algols (H Algols),
Helium Cataclysmic Variables (He CV), and Helium Algols (He Algols),
based on the evolutionary stage of the secondary,
method of mass transfer, and composition of the
secondary's envelope (\cite{branch95}; \cite{ruiz96}).
The secondary can be a low-mass main sequence, subgiant, or red giant star
that slowly loses its envelope to the primary by Roche lobe overflow
for close binaries or by stellar winds for wider binaries.
The mass transferred can be hydrogen or helium in cases where the hydrogen
layer has already been lost by earlier phases of mass transfer.

In the H CV scenario, a CO white dwarf,
left in a close binary orbit by an earlier episode of common
envelope evolution in its AGB phase, accretes hydrogen by Roche lobe overflow
from a lower-mass main sequence secondary.
The mass transfer is maintained by magnetic braking and
has a characteristic rate of \.{M} $>1.0 \times 10^{-8}$ 
\Msunperyear~ (\cite{verbunt81}).
Not all H CV systems end as Type Ia supernova progenitors.
Only the largest white dwarfs
with the highest mass transfer rates, high enough to suppress hydrogen flashes,
can accrete enough mass to reach the \Ch mass.
Livio \& Truran (1992) associated these binaries with recurrent nova systems.
When the \Ch mass is reached, the white dwarf explodes as a
Type Ia supernova, but only if it can avoid accretion-induced collapse (\cite{nomoto90}).
Della Valle \& Livio (1994) suggested these binaries could be responsible
for Type Ia events in late-type galaxies while another progenitor
from an older stellar population could be responsible for
Type Ia events in early-type galaxies.

The H CVL systems have an evolutionary path like that of the H CV systems.
After an episode of common envelope evolution, a white dwarf is
left in a close binary orbit with a slightly higher-mass 
main sequence or subgiant secondary.
Because of the mass ratio, the mass transfer is unstable
and proceeds on a thermal timescale. This higher
mass transfer rate (\.{M} $\sim 10^{-6}$ \Msunperyear)
is responsible for the steady burning of
hydrogen to helium on the surface of the white dwarf.
Eventually, the \Ch mass is reached and the star explodes
(\cite{rappaport94}; \cite{hachisu96}; see also \cite{hachisu99b}).

We use the H Algol and Symbiotic classification, following the path 
of Branch \etal (1995), as a general term to describe binaries in which 
the donor is a red giant that slowly loses mass to the white dwarf.
In these scenarios, after a common envelope phase,
a CO white dwarf is left in a binary orbit with  a main sequence star.
In the H Algol scenario, the main sequence secondary
evolves all the way to the red giant phase before filling its Roche lobe 
for the first time (\cite{whelan73}; \cite{iben84}). Mass transfer
to the white dwarf is by Roche lobe overflow, 
like the other Type Ia progenitor scenarios.
In the symbiotic scenario, the red giant is too distant to fill
its Roche lobe in the course of its evolution. Mass loss will 
occur by stellar winds.
Many evolutionary paths can lead to the symbiotic scenario, 
the simplest being a wide binary whose stars
evolve independently. Other possible paths include
prior phases of common envelope evolution or
conservative mass transfer during the primary's evolution,
which leaves the binary separation too wide for the secondary to ever fill
its Roche lobe (\cite{yung95}). 
Until recently, symbiotics were considered as viable Type Ia candidates
only for \subCh explosions (\cite{kenyon93}; \cite{yung95}).
However, recently Hachisu \etal (1999a) have proposed that symbiotic systems 
with \Ch explosions can account for the Type Ia rate.

The He CV and He Algol classes are analogous to the
H CV and H Algol classes, the primary difference being
that the mass transferred is helium instead of hydrogen.
This complicated evolutionary path begins with
the creation of a white dwarf in a close binary orbit,
following a phase of common envelope evolution
with a main sequence star, as in the previous scenarios.
The secondary evolves though the red giant phase. However, unlike
the earlier scenarios, the secondary catastrophically
loses its hydrogen envelope in another common envelope phase.
The secondary, now a helium star, continues to evolve.
For the He CV scenarios, mass transfer begins when the
helium star goes into Roche lobe overflow. Stable mass transfer ($q<1$) is
maintained by gravitational wave radiation and proceeds
on a nuclear timescale. 
For these binaries, helium star masses are typically in the range of
$0.3-1.0$ \Msun and white dwarf masses are in the range of $0.6-1.0$ \Msunn.
These are Type Ia supernova \Ch and \subCh explosions (\cite{iben91}).
For the He Algol scenarios, mass transfer begins
during the secondary's expansion to a helium giant.
Mass transfer is quasi-stationary (M$\sim 10^{-6}$ \Msunperyear) with 
secondary masses in the range of $0.75-2.3$ \Msun and with white dwarf masses
in the range of $0.9-1.4$ \Msunn.  The He Algol systems are candidates for
Type Ia \Ch explosions (\cite{iben94}). Branch \etal (1995) finds that
the realization frequencies for these scenarios are $10^{-4}-10^{-5}$ yr$^{-1}$
for young populations and are negligible for old populations.

In this paper, we focus on exploring the supernova-secondary
interaction for the four hydrogen-rich scenarios in Table \ref{singlemodels}.
To simulate each impact, we select a representative binary system
from the literature for each of the four scenarios and estimate 
the binary separation and the companion's mass at the time of explosion.
We use a stellar model in the simulation that is reasonably 
close in mass and evolutionary stage. We use either SNIa W7 (\cite{nomoto84})
to represent a \Ch explosion or SNIa Hedt (\cite{woosley94}, hedtb11) to represent
a \subCh explosion, depending on which is more likely for
each binary scenario.
The secondary models are listed in Table \ref{secondarymodels} 
with the masses and radii provided. Information on the supernova ejecta 
models is given in Table \ref{ejectamodels}.

To represent the H CV class, we construct a prototype binary, based on the
Type Ia supernova progenitor candidates of Livio \& Truran (1992), consisting
of a $1.0$ \Msun main sequence secondary that lost $0.1$ \Msun 
to a massive $1.3$ \Msun white dwarf primary, allowing it to reach
the \Ch mass.  Assuming that the binary is in Roche lobe overflow
at the time of explosion and that the secondary just
fills its Roche lobe, from Eggleton's relation (\cite{eggleton83})
for $q = 1.0/1.4$
we estimate a binary separation of $\sim 3$R.
For this simulation, we represent the secondary by a $1.0$ \Msun 
solar model (\cite{chaboyer98}), 
the exploded white dwarf by the SNIa W7 (\cite{nomoto84}) ejecta profile, 
and employ a binary separation of $3$R. 

We use the work of Li \& van den Heuvel (1997) on supersoft X-ray sources as Type Ia
supernova progenitor candidates to construct sample scenarios
for both the H CVL and H Algol classes, one for each candidate
proposed by Li \& van den Heuvel (1997). Although H CVL binary systems have
been included in the past as part of population synthesis studies
(\cite{rappaport94}), they are now the subject of renewed interest
due to the suggestion by Hachisu \etal (1996) that a 
white dwarf fed by an optically thick wind can experience a much higher 
mass transfer rate than previously thought.
Hachisu \etal (1996) have suggested that these binary
systems are the progenitors of some supersoft X-ray
sources, as well as potential Type Ia supernova
progenitors, although the realization frequency
is in dispute (\cite{yung98}).

As prototypes for our H CVL and H Algol classes,
we select the two Li \& van den Heuvel (1997) scenarios: the first scenario
with a $2.5$ \Msun subgiant companion and the second with a
$1.0$ \Msun red giant companion, prior to mass transfer.
We represent the subgiant companion just prior
to mass transfer with a $2.1$ \Msun subgiant star having 
a radius of $2.3$ \Rsun (\cite{chaboyer98}).
To simulate the effect of binary mass transfer, which was not
included when the star was evolved,
we  adjust the entropy profile to decrease the mass 
and radius of our secondary to $1.13$ \Msun and $1.7$ \Rsun, respectively,
which reproduces the mass and radius of the secondary at the time 
of explosion as estimated from Li \& van den Heuvel (1997).
We do not attempt to reproduce the effect of the mass transfer
on the red giant secondary. We represent the red giant companion
with a $0.98$ \Msun red giant with a radius of $170$ \Rsun 
(\cite{chaboyer98}).
Because we can not include binary mass transfer in
the stellar evolution code, our stellar models
can not be exact  recreations
of the secondaries used in the binary evolution
calculations of Li \& van den Heuvel (1997).
However, as we shall see,
the stripped fraction and the velocity
distribution of the stripped material
are fairly robust and predictable.

To represent the Symbiotic (SS) class, we construct
a prototype system based on the work of Yungelson \etal (1995),
who concluded that symbiotic systems could be
the progenitors of up to $1/3$ of Type Ia events
in young and intermediate age (t $\lesssim 6 \times10^{9}$ yr)
stellar populations.
We select the same $0.98$ \Msun red giant secondary as in the
H Algol case, but 
to ensure that mass transfer takes place by stellar winds
we increase the binary separation so that the
red giant could not have filled its Roche lobe.
Because these systems are good candidates for
Type Ia \subCh explosions, we select the ejecta model SNIa Hedt 
(hedtb11, \cite{woosley94})
to represent the exploded white dwarf. 

Table \ref{sims} lists the key simulations performed for this paper,
with the stellar model, supernova ejecta model, and binary separation
that were used in each simulation. The four simulations, representing
the four hydrogen-rich classes of single-degenerate progenitor models,
are labeled after their respective class as HCV, HCVL, HALGOLa, and
SYMB.  Additional calculations were performed (HCVa, HCVb, HCVc, and HCVd) 
to gauge the trends with separation in the mass stripped, the kick, and the velocity
distribution of the stripped hydrogen.
We include a simulation using a
$2.1$ \Msun subgiant (HCVLa) secondary, which we compare
to the HCVL simulation, whose secondary suffered substantial 
mass loss prior to the supernova explosion.

Our choice of reference binary scenarios is motivated 
in part by the likelihood that a particular progenitor
can be responsible for a significant fraction of Type Ia 
events, and in part by the desire to cover the range of
parameter space likely to be encountered by any Type Ia explosion
in a single-degenerate binary system. We expect that the four binary scenarios
HCV, HCVL, HALGOLa, and SYMB will cover the basic range of 
secondary stars, supernova explosion models,
and binary separations expected for Type Ia supernova progenitors 
with a minimum of overlap. 
For example, the HCV, HCVL, and the HALGOLa simulations represent
low-mass main sequence, subgiant, and red giant companions in 
Roche lobe overflow,
in order to cover the range of compactness in the the evolution
of a low-mass star. The mass profile of each secondary model is shown 
in Figure \ref{m}, and the binding energy profile is shown in Figure \ref{be}.
The kinetic energy of the supernova blast
is roughly $11$, $10$, $2300$, and $1300$ times the
binding energy of the secondaries for the HCV, HCVL, HALGOLa, and SYMB
simulations, the difference being the
large difference in binding energy between the tightly-bound 
main sequence and subgiants and the loosely-bound envelope of 
the red giant. To explore the role of the supernova debris profiles,
we employ two Type Ia models
(see Table \ref{ejectamodels}): W7, a standard 
\Ch explosion model, and Hedt, a \subCh model with
$78\%$ of the momentum of W7. The HALGOLa and SYMB simulations
represent the red giant companion with the same secondary model,
but the HALGOLa simulation represents the supernova ejecta with W7.
The SYMB simulation represents the ejecta with Hedt at a slightly 
greater binary separation.
Because the mass transfer is wind-driven,
the solid angle subtended by the red giant in the SYMB simulation
is therefore slightly smaller.
The solid angle subtended determines the fraction of the supernova
momentum incident on the secondary. For any low-mass companion in Roche lobe overflow 
the fraction of the sky subtended is typically $(1/4)(R/a)^2 \sim 1/36$ 
because a$\sim 3$R, independent of the companion's stellar type.

\section{The Main Sequence Companion}
\label{sec:mainsequence}

In this section, we describe the impact of the
blast wave on a main sequence star, our HCV scenario.
As in the subgiant and red giant
cases, we neglect any changes to the structure
of the star from the Roche potential.
Likewise, we do not add an accretion disk,
magnetic field, or any other
intrinsic features of cataclysmic variables
which would introduce complications.
For this simulation,
we use a $300 (\rho) \times 575 (z)$ cylindrical grid
with the $1.0$ \Msun main sequence secondary,
centered on the origin.
Approximately 
$161 (\rho)\times 322 (z)$ zones are allocated to the secondary.
The grid extends
$6$ stellar radii in the $\rho$ direction, 
$12$ stellar radii in the downstream (negative $z$) direction 
and $2$ stellar radii in the upstream (positive $z$) direction. 
The white dwarf primary explodes just off the grid in the
upstream direction, and the supernova debris flows through the
upper boundary onto the grid by time-dependent
boundary conditions.
The density and velocity, which characterize the debris structure,
energy, and momentum, are scaled with time: the Eulerian velocity as $t^{-1}$
and the Lagrangian density as $t^{-3}$.

The HCV was run for $2.0\times10^{4}$ seconds,
slightly more than $7$ of the companion's sound crossing times 
and much longer than the $1.4\times10^{3}$ seconds that
it takes for the trailing edge of the supernova to sweep
past the back of the companion.
The length of the calculation is chosen
to be sufficiently long, and the size of the grid sufficiently large,
that we can follow the progress of the shock-heated
material ejected from the companion
in both velocity and solid angle
and can construct a distribution for each.
In addition, the long simulation time allows 
the companion star to begin to
reestablish hydrostatic equilibrium. 
Although we can not follow the companion's Kelvin-Helmholtz evolution,
its post-impact structure may
provide us with some insight into the possible consequences 
of the impact for the companion's future evolution.

For clarity, this section is subdivided into distinct topics.
In \S \ref{sec:hydroms}, we describe the debris structure of the impact,
and in \S \ref{sec:hburn} we consider the consequences of the
impact for hydrogen burning.
In \S \ref{sec:stripms}, we present and describe the stripped mass for
this scenario. 
In \S \ref{sec:sncontamination}, we qualitatively discuss the possibility that
the main sequence companion may accrete supernova material
and contaminate its envelope with heavy metals.
In \S \ref{sec:distms}, we discuss the
velocity and solid angle distributions
of the stripped companion material and possible observational 
implications.
In \S \ref{sec:kickms}, we discuss the
kick received by what remains of the companion.

\subsection{Hydrodynamics of the Impact on the Main Sequence Companion}
\label{sec:hydroms}

\newcommand{\frameA}{\ref{ms1}a~}
\newcommand{\frameB}{\ref{ms1}b~}
\newcommand{\frameC}{\ref{ms1}c~}
\newcommand{\frameD}{\ref{ms1}d~}
\newcommand{\frameDD}{\ref{ms1}d}

\newcommand{\frameE}{\ref{ms2}a~}
\newcommand{\frameEE}{\ref{ms2}a}
\newcommand{\frameF}{\ref{ms2}b~}
\newcommand{\frameFF}{\ref{ms2}b}
\newcommand{\frameG}{\ref{ms2}c~}
\newcommand{\frameGG}{\ref{ms2}c}
\newcommand{\frameH}{\ref{ms2}d~}
\newcommand{\frameHH}{\ref{ms2}d}

\newcommand{\frameCB}{\ref{ms3}a~}
\newcommand{\frameCBB}{\ref{ms3}a}
\newcommand{\frameCC}{\ref{ms3}b~}
\newcommand{\frameCCC}{\ref{ms3}b}
\newcommand{\frameCD}{\ref{ms3}c~}
\newcommand{\frameCDD}{\ref{ms3}c}
\newcommand{\frameCE}{\ref{ms3}d~}
\newcommand{\frameCEE}{\ref{ms3}d}
\newcommand{\frameCF}{\ref{ms3}e~}
\newcommand{\frameCFF}{\ref{ms3}e}
\newcommand{\frameCG}{\ref{ms3}f~}
\newcommand{\frameCGG}{\ref{ms3}f}

The impact of a blast wave on a main sequence
star begins a sequence of events common to all
supernova-companion interactions.
The initial impact of the supernova shell on the surface 
of the star drives a shock into the stellar envelope.
A reverse shock propagates back into the ejecta.
A contact discontinuity
marks the interface between the supernova ejecta and the shocked
stellar material. The shock propagating through the
envelope decelerates as it runs up the steep
density gradient of the star. After the shock passes
through the stellar core, it accelerates down the 
decreasing gradient. The shock front, highly curved
because of the density gradients, converges in the
back of the star. 
The reverse shock develops into a bow shock 
around the companion star that smoothly deflects the rest
of the incoming supernova debris around the companion. 
After most of the supernova debris has passed by, the outer layers
of the stellar envelope, which have been shock-heated to such an extent
that the new speed of sound exceeds the companion's
escape velocity, are ejected, often corrugated by \kh and \rt instabilities.
The stripped material flows slowly away from the
star, embedded within the inner layer of the supernova ejecta.
The stellar core expands and cools in response to the initial compression 
by the shock.
The star now has an extended, asymmetrical envelope and
begins to pulsate to bring the envelope back into
hydrostatic equilibrium.
Figures \ref{ms1} - \ref{ms2} illustrate the 
sequence of the interaction in a series of 2-D images
from the initial impact
to the reestablishment of hydrostatic equilibrium.
Figures \ref{ms3} illustrate the mass stripping of
the companion in a series of 2-D cartoon images, with
each color indicating the dominant element in that 
region. 

Figure \frameA shows 
the companion centered at the origin of the cylindrical
grid at the beginning of the simulation.
Figure \frameB shows the initial impact of the blast at
$\sim 100$ seconds after the explosion.
The supernova ejecta can be seen flowing onto the
grid, sweeping up ``circumstellar'' material
as the leading edge begins to flow around the secondary star.
There is a region of high pressure at the site of the impact on
the companion's surface where a shock is being driven into the
star, and a bow shock is beginning to develop that will 
divert the flow of the supernova ejecta around the
companion. The back of the companion is still undisturbed.

The density variations in the ejecta that are visible as
spherical rings in Figure \frameB 
can be directly identified with the 
density variations in Figure \ref{SN} by their velocity.
For example, the first strong ring striking the 
surface of the companion star in Figure \frameB has a
velocity of  $1.59\times10^{4}$ \kmsec and can be
identified with the first sharp density spike in 
Figure {\ref{SN}. The second ring has a 
velocity of $1.36\times10^{4}$ \kmsec and can be
identified with the second density spike in 
Figure {\ref{SN}. These features are associated with
changes in the composition of the ejecta: the transitions
between the ``oxygen'', ``silicon'', and ``iron'' layers.
(see Figure \frameCBB).

Figure \frameC shows the impact $179$ seconds after
the explosion. The tail shock and bow shock 
dominate this image. By this time, the collision of the 
supernova ejecta with the ``circumstellar''
medium has created a curved shock front which
has converged along the downstream axis after flowing
around the companion.
The shock front reflects at the axis and begins to move outward.
The tail shock is visible in Figure \frameC as
the flared vertical discontinuity along the downstream axis.
The upstream axis is dominated by the bow shock which wraps
smoothly around the top of the companion, but becomes
distorted farther down as a consequence of the impact of
the density variations of the ejecta.
The material between the bow shock and the shocked surface 
of the star is primarily supernova debris mixed with ``circumstellar'' medium. 
The pressure in this region now rivals the pressure in the
interior of the star (see Figure \ref{press1_ms}).

At $329$ seconds (Figure \frameDD), the shock front passing 
through the companion has a strong curvature
caused by the density gradient of the star.
The center of the shock decelerates as it encounters the rising 
density near the stellar core, while the wing of the
shock accelerates through the back half of the companion
where the density is decreasing.
The shock-heated envelope of the companion starts to 
suffer mass loss; escaping material is now visible in the flow 
that passes around the star behind the bow shock. \kh and \rt
instabilities in
the shocked envelope are just barely visible in this image.

The ejected companion material is confined to a torus
around the back of the companion - with the inside of the torus
filled with ``circumstellar'' material that has been compressed
behind the companion. There, 
the flow is almost directly parallel to the downstream axis
(see Figure \frameCDD).
This material is the first to be
ejected from the star and is mixed with the oxygen layer
of the supernova debris. This
will form the high-velocity tail of the velocity distribution
of the stripped companion material.
In Figure \frameCDD,
the iron shell has expanded to enclose the star,
the silicon shell is just below the back of the star, and
the oxygen shell is everywhere exterior to the silicon shell.
More than $90\%$ of the supernova's mass and momentum that will
collide with the companion is now represented on the grid.

Figures \frameE and \frameF show the interaction at
$529$ and $755$ seconds after the explosion
(see also Figures \frameCE and \frameCFF). By $529$ seconds,
the collision is effectively finished. Only a negligible mass and 
momentum will collide with the companion after this time.
The shock within the companion has just passed through the
stellar core and will begin to accelerate down the
density gradient of the back of the companion. 
\kh and \rt instabilities, the fingers of swirling material
in the shocked envelope, are very visible by this time.
As yet, only $0.07$ \Msun has been ejected. 
By $755$ seconds, the shock within the companion has converged
along the downstream axis. 
In subsequent frames, a reflection shock can be seen progressing
back through the remnant of the companion.

Figure \frameG shows the end of the mass stripping phase.
By $2027$ seconds, the star has expanded and pushed
the bow shock back towards the site of the explosion.
About $93\%$ of the companion's mass that will be ejected has 
been ejected by this time. 
The velocity field around the star is now more spherical.
This can be inferred from Figure \frameCG by the change in
the distribution of the stellar hydrogen, 
which is no longer confined along the downstream axis.
About $1000$ seconds later,
the first in a series of shocks is driven into the envelope 
marking the beginning of a global stellar pulsation phase.

Figure \frameH shows the remnant after three oscillations,
$2\times10^{4}$ seconds after the supernova explosion.
The star is almost spherical again, but
it is surrounded by a complicated system of shocks. The shock
structures can be qualitatively explained in terms of the
overtaking of each shock front by 
the subsequent shock front. Add to this mixture of shocks
the reflected shock from the original shock through the
companion, and the structure becomes difficult to disentangle.

Exactly what happens to the structure of the companion as 
the shock is driven through the star? 
To reliably follow the companion from the impact to
the reestablishment of hydrostatic equilibrium,
we ran a slightly different simulation, with the same binary scenario,
but with a uniform grid and the gravity solver centered on the
center of mass at each timestep, which is
more suitable for long simulation times.
Figures \ref{press1_ms} - \ref{press3_ms}
show the evolution of the secondary's pressure profile from 
the beginning to the end of that simulation.
Figure \ref{press1_ms} shows the shock climbing the density gradient
towards the stellar core.
At $\sim 400$ seconds, the shock passes through the stellar core
and begins to descend down the density gradient.
Also, as a result of the impact, a reverse shock
is formed at the interface between the star and supernova that 
propagates back into the supernova material.
Eventually, the reverse shock becomes the bow shock.

As Taam \& Fryxell (1984) discussed in their 
work on impacts on low-mass main sequence secondaries,
the ram pressure of the ejecta is roughly equal to the 
central pressure of the main sequence secondary.
The typical ram pressure of the ejecta is
$P = \rho v^2 \approx M_{ej} v^{2} / (4 \pi a^{3}/3)$,
(N.B., a $\sim 3 R$ when it hits a low-mass secondary in Roche lobe overflow).
For our main sequence simulation this is $\sim 10^{17}$
\dynescmcm, very close to the central pressure of
$2.37\times10^{17}$ \dynescmcm.
The ram pressure of the ejecta is close to the central pressure
of the secondary only in the simulation with the main sequence companion.
The more evolved subgiant and red giant have
larger radii, which increases the binary separation and, thereby,
decreases the incoming density and the ram pressure at first impact.
For our subgiant and red giant secondaries, the
expected ram pressures are roughly $10^{16}$ and $10^{10}$
\dynescmcm, respectively,
which is too low to significantly affect the
subgiant's core by $2$ orders of magnitude and
the red giant's core by $7$ orders of magnitude.
In our HCV calculation, we find that the central pressure changes by a
factor of only $2.5$ as the shock passes through the core,
which makes it a weak shock.
The central density rises by only a factor of $\sim 1.7$, the
temperature by $\sim 1.5$ and the entropy by 
$2.5$ k$_{B}^{-1}$ baryon$^{-1}$.

After the initial pressure rise, the star expands and 
the central pressure begins to drop. 
The shock accelerates down the density gradient in the back of the star until 
at $\sim 700$ seconds it passes through the limb of the star and
converges on the downstream axis.
Figure \ref{press2_ms} illustrates the dramatic pressure rise at the
back of the star as the curved shock fronts converge.
The high back-pressure exerts a force on the secondary which 
decreases the kick received from the impact.
As part of the shock-heated envelope escapes from the star,
the back-pressure drops,
until after approximately $2000$ seconds, the
mass stripping phase is complete and the remainder of the 
secondary reaches its terminal velocity.
At this stage, the star is slowly expanding. However, the
passage of the shock and the ejection of part of the
envelope begin a phase of radial pulsations during which
shocks are sent out through the extended envelope.
It is during this phase that the post-impact secondary
recovers hydrostatic equilibrium.

Figure \ref{press3_ms} shows the pressure profile from the
end of the mass stripping phase to the reestablishment of
hydrostatic equilibrium.
As slow moving material begins to settle back onto the
star, a bounce occurs ($\sim 4000$ seconds) and a shock is driven into
the outer envelope.
Every $3000-4000$ seconds, roughly a sound
crossing time, another
pulsation with a smaller amplitude is seen moving through 
the extended envelope.
Despite the dramatic appearance of the shocks
in the extended envelope, little or no mass ejection takes place
as a result.
As shown in Figure \ref{DTts98}, the stellar core also pulsates,
with a period of $\sim 2600 $ seconds.
In the last pressure profile shown in Figure \ref{press3_ms},
the central pressure has stabilized. 

The changes in the central pressure, density, temperature,
and entropy are listed in Table \ref{postimpactchart}.
All quantities but the entropy, which increases by shock-heating,
have dropped significantly due to the expansion.
Even after the interior of the star stops expanding and 
the infalling material compresses the core, the central
pressure and density are not restored to their original values.
Although such pulsations were not noted in earlier numerical
simulations, probably due to shorter simulation times,
Taam \& Fryxell (1984) did find that some of their low-mass main
sequence companions suffered an abrupt rise in central density
followed by a severe drop as the star began to expand.
The density and entropy profiles of the pre- and post-impact
companion are shown in Figures \ref{Dens1} and \ref{S1}.
The radius of the remnant star, which we define as
the radius that encloses about $90 \%$ of the remnant's mass,
is approximately $7.4\times10^{5}$ km, 
a modest increase of less than $10\%$.
The remaining $10\%$ of the mass forms a low-density extended envelope 
out to $6.7\times10^{6}$ km ($\sim 10$ \Rsunn) with a composition
similar to that of the original stellar envelope.

\subsection{Quenching Hydrogen Burning in the Stellar Core}
\label{sec:hburn}

Do the abrupt rise and fall of the central temperature
and density have any unique consequences for the energy
generation rate and the luminosity of the companion during the
impact? 
We expect that the energy generation rate, with its
extreme temperature sensitivity, should rise and fall
dramatically as the stellar core heats and cools.
Figure \ref{DTts98} illustrates the rise and fall of
the central temperature and density.
The oscillations that the secondary experiences as hydrostatic
equilibrium is reestablished are visible in both profiles.
For a quick estimate of the energy generation rate,
we calculate the hydrogen burning rate via
the three pp chains and the CNO cycle (\cite{clayton83}; \cite{kippenhahn91})
along a 1-D cut through the densest (and hottest) part of the
companion. As can be seen in Figure \ref{NucTs98},
the energy generation rate shoots up by a factor of $1000$ as a result
of shock-heating, then drops precipitously as the
star expands. 
Because the compression phase is so brief,
despite the extremely high energy generation rate,
only a small amount of additional energy is deposited in the
center of the star.
For the main sequence star an additional $\sim 10^{38}$ ergs
has been deposited, which is extremely small in comparison
with the binding energy of the star ($\sim 10^{48}$ ergs).
However, because the timescale of the compression phase is close to
the sound crossing time of the stellar core, the star may
be able to expand and cool. Hence, $10^{38}$ ergs is an upper limit.
By the end of the simulation, the energy generation
rate is only $0.07$ \Lsunn. The hydrogen burning has
been quenched by the impact. The implications of this
are discussed in \S \ref{sec:postimpact}.

\subsection{The Stripped Mass of the Main Sequence Companion}
\label{sec:stripms}

The amount of mass stripped from the secondaries is shown 
in Table \ref{stripchart}.
The third column lists the stripped mass
for each scenario based on the numerical calculation and the fourth
column lists the stripped mass based on the analytic estimation 
of Wheeler \etal (1975). 
To determine the amount of material ejected
by the impact and the passage of the shock through the
star, we compare the velocity of the companion's material 
within each zone with its local escape velocity 
$\sqrt{ 2 \phi }$, $\phi$ being the gravitational potential.
If its velocity exceeds the escape velocity, we consider the mass 
unbound. If not, we consider the mass
bound to the star, only displaced, 
although, because of the Eulerian nature of the code, we are not able to
follow the calculation until the displaced stellar material
falls back onto the companion. 
When any stellar material reaches the edge of
the grid, we assume that if it is unbound at that time it
will remain unbound.
By the end of the simulation,
$0.15$ \Msun of the main sequence stars's envelope is unbound, and
all but $ 2.4\% $ of this material has left the grid.
Although we run the simulation for $2\times10^{4}$ seconds,
$93\%$ of the stripped mass is ejected in the first $2\times10^{3}$ seconds,
which is longer than the $1.4\times10^{3}$ seconds for the supernova
to pass by the secondary.

In Table \ref{stripchart} we compare the numerical stripped mass with
an analytic estimate. The analytic estimate is
within $10\%$ of the numerical estimate, surprisingly close,
although that changes as the binary separation increases.
A detailed comparison of the numerical results with the analytic estimates 
is in \S \ref{sec:separation}.
For now, we note that the analytic estimates do provide ballpark estimates of the
stripped mass for the main sequence case.
Because we use a smaller $a/R$ ratio, consistent with Roche lobe overflow,
our $0.15$ \Msun of stripped material is substantially higher than
the $0.013 - 0.052$ \Msun reported by Fryxell \& Arnett (1981).
However, we can compare our HCVc simulation to Fryxell \& Arnett's CASE C, 
which has a similar $a/R$ ratio, by scaling our stripped mass fraction
by the new incident supernova momentum.
After the scaling, our $0.022$ \Msun of stripped material
increases to $0.057$ \Msun which is not inconsistent with the
$0.052$ \Msun they report.

\begin{sloppypar}
\subsection{Contamination of the Main Sequence Companion with Supernova Debris}
\label{sec:sncontamination}
\end{sloppypar}

It is possible that some supernova material may be 
mixed into the envelope of the secondary star, either as a direct 
consequence of the impact or as fallback.
The enhancement of heavy elements in the atmosphere of the
secondary star may be observable $10^{3}-10^{4}$ years 
(the \kh timescale) after the impact
or might change the course of the secondary's post-impact \kh evolution
by increasing the opacity of the stellar envelope.
The question of how much supernova material is likely to be
accreted by the secondary star as a direct result of the explosion
is of particular interest in connection with
the overabundance of $\alpha$-elements observed in the atmosphere of the
secondary star in Nova Scorpii 1994 (GRO J1655-40) 
(\cite{israelian99}). This binary X-ray source is believed to consist of
a black hole primary in close orbit with a subgiant
secondary. Israelian \etal (1999) have interpreted the
anomalous abundances in the subgiant's atmosphere as contamination by
fresh nucleosynthesis products ejected during a Type II supernova explosion
that might have accompanied the formation of the black hole.

Our calculations imply
that the stellar remnant is not contaminated by
the supernova material as a direct result of the impact.
At the end of our main sequence simulation, no significant amount of 
supernova material is within the original radius of the secondary. 
However, a trace amount ($1.3\times10^{-4}$ \Msunn) of supernova
material, predominately iron-rich $(\ge 90\%)$ is bound
by the secondary star, although the only supernova material
in the immediate vicinity of the secondary star at the end 
of the main sequence simulation is physically located exterior 
to the bow shock.
It is possible that the secondary may accrete this low-velocity
iron and any additional iron mixed with the marginally-bound 
companion material when
the displaced envelope, which is embedded within the iron layer
of the supernova ejecta, begins to fall back.
In addition, the slow trailing edge of the supernova ejecta
could act as a source of iron-group elements long after the
impact. 

We can make a ballpark estimate of the amount of supernova debris that can
contaminate the secondary star by finding the amount of mass of the
supernova ejecta which can not escape the gravitational potential of the
secondary. A typical escape velocity for a $1.0$ \Msun main sequence 
star is $\sim 600$ \kmsecc. From the supernova ejecta model W7,
we find that $\sim 1.5\times10^{-3}$ \Msun has a velocity lower than the escape
velocity; we can use this as a crude and conservative upper limit for the 
degree of contamination. 

\subsection{Distribution in Velocity and Angle of the Material Stripped from the
           Main Sequence Companion}
            
\label{sec:distms}

One of our key results is the velocity distribution of the stripped
hydrogen material, shown in Figure \ref{dmdv_bw2}a,
for the HCV simulation. Also plotted in this
figure is the velocity distribution of W7, the supernova
ejecta model. The first feature to catch the eye is the
clear separation in velocity space between the stripped material 
and the supernova ejecta. 
The velocity at the half-mass point in the velocity profile of
the main sequence secondary is
$823$ \kmsecc, much less than the half-mass
velocity of the supernova ejecta of $7836$ \kmsecc.
Each half-mass velocity is marked with a vertical line and an
arrow in Figure \ref{dmdv_bw2}a.
The high-velocity tail shows three sharp spikes, which
we identify with the three sharp density peaks
at the composition transitions in SN W7. 
The composition groups for both the stripped material and
supernova are shown; from this, it is clear that the
bulk of the stripped material, predominately hydrogen,
is embedded within the iron layer of the supernova ejecta.
This result reinforces the belief that contaminating
hydrogen and helium are most likely to be observed as late-time
emission when the photosphere of the supernova moves
to low-velocities, revealing the core.
However, a high-velocity tail does exist, implying that
a trace of companion hydrogen is swept up in the oxygen and
silicon layers of the supernova ejecta.
Given the upper limits on the hydrogen
abundance from Type Ia observations near maximum
light, this may provide a criterion for
discriminating between Type Ia progenitor scenarios.

In Figure \ref{dmdv_bw2}a
we correct the low-velocity tail by allowing the
stripped material to escape to infinity, free of the
secondary's gravitational potential.
If we assume the stripped material flows off the grid at
high-Mach numbers (typically M$ \sim 5 - 30$), 
then the specific enthalpy
\begin{math}
w = c_{s}^{2} / (\gamma -1) 
\end{math}
is small in comparison with the
kinetic energy, and we can neglect the enthalpy in
Bernouilli's Equation 
\begin{math}
w + 1/2 v^{2} + \phi = C,
\end{math}
where $C$ is a constant along a streamline.
Therefore, we can estimate
the velocity of the flow at
infinity as
\begin{math}
 v_{\infty}^{2} = v^{2} - v_{esc}^{2}.
\end{math}
In Figures \ref{dmdv_bw2}
we have subtracted the escape velocity in quadrature.
The subtraction in quadrature affects only the
low-velocity tail.

Figures \ref{dmds_ts85_ts88} and \ref{polar_ts85_ts88} show the
spatial distribution of the stripped mass.
Figure \ref{dmds_ts85_ts88} illustrates the distribution
of the stripped material in solid angle
(dM/d$\Omega$) in the HCV simulation,  expressed in terms of the angle, 
$\alpha$, from the downstream axis.
The half-mass point of the HCV distribution is $\sim 43^\circ$, so
about half of the stripped material is confined in
a cone within about $\sim 43^\circ$ of the downstream axis
and the other half extends out to $\sim 90^\circ$.
The sudden rise along the axis for the HCV distribution 
is a consequence of the use of 
the differential element d$\Omega = 2 \pi \sin \alpha d\alpha$ when
$\alpha \rightarrow 0$.
Figure \ref{polar_ts85_ts88} illustrates the same spatial
distribution as a polar plot. 
The polar angle, $\alpha$, once again represents the angle from the downstream 
axis, and the radial coordinate represents the magnitude of $dM/d\alpha$.
The profile clearly shows that the preferred direction is along
the $43^\circ$ line. 
In addition, the majority of the mass is exterior to the geometrical shadow,
the shadow that the secondary would cast from the site of the
explosion of the primary. 
Because of the compression of the ``circumstellar'' material along the
downstream axis as the supernova sweeps around the star,
very little stripped material is directly behind the secondary.
In a 3-D calculation instabilities along the downstream axis
are likely to mix the compressed ``circumstellar'' material with
the stripped stellar material, filling in this region.

As shown in Figure \ref{dmds_ts85_ts88},
the collision of the supernova debris with the secondary will create 
a hole in the debris structure. The highest velocity debris will flow
around the secondary, creating a cut-out region. The lower velocity
material will flow around the bow shock, creating a wider hole
in the inner ejecta.  
For the HCV simulation, the supernova debris will have a hole
of $\sim 31^{\circ}$ in the outer ejecta, which will widen
to $\sim40^{\circ}$ in the inner ejecta. 
This corresponds to $7\%$ and $12\%$ of the surface of the
supernova ejecta. 
Due to the supersonic nature of the flow, 
the hole will not close with time. 
Hence, the asymmetry in the supernova remnant may indicate the
presence of a secondary star.

If the stripped hydrogen were ejected spherically and
uniformly mixed with the (unperturbed) supernova ejecta,
then the mass fraction of hydrogen would be
$0.6$ to $0.7$, almost solar, in the low-velocity region 
($\lesssim 10^{3}$ \kmsecc).
The hydrogen mass fraction, X(H), is shown 
as a function of supernova mass and velocity 
in Figures \ref{Mts85} and \ref{Vts85}.
The hydrogen mass fraction falls abruptly 
as the velocity increases.
At $10^{4}$ \kmsecc, which is near the oxygen and
silicon layers, X(H)$\sim 3\times10^{-3}$. 
Of course, we know from the numerical simulation that the
ejected material contaminates a wide solid angle behind
the secondary. For the HCV simulation, $90\%$ of the
stripped hydrogen lies within $65.7^{\circ}$ of the
downstream axis
which corresponds to a fractional solid angle of $29\%$.
Figures \ref{Mts85} and \ref{Vts85} show the expected 
mass fraction within the contaminated solid angle.
As expected, the mass fraction rises dramatically.

\subsection{The Kick Received by the Main Sequence Companion}
\label{sec:kickms}

Table \ref{kickchart} lists the kicks given to the
remnant companions in the main sequence and subgiant scenarios. 
We note that the red giant core, because of its extremely small
size, receives a negligible kick.
We calculate the kick from the
supernova impact by finding the terminal velocity of the center of 
mass of the unstripped stellar material that remains on the grid.
Although some material still bound to the
star can flow off the grid, it is a negligible fraction
($\Delta M / M \le 10^{-2}$), since the star
reaches its terminal velocity very early in the simulation.
For the main sequence simulation HCV, the companion
receives a kick of $\sim86$ \kmsecc. Although substantial,
this is still much less than its $227$ \kmsec orbital velocity.

In the HCV simulation
the terminal velocity is reached by $5000$ seconds after
the explosion. The secondary receives a very strong kick ($> 100$ \kmsecc)
in the first $\sim 500$ seconds as the shock passes through
the center of the companion. As the shock converges and reflects 
in the back half of the companion ($\sim 700$ seconds), 
the pressure rises dramatically, and 
the companion, now elongated, decelerates. By $1500$ seconds,
part of the shock-heated envelope has escaped the star and
the over-pressure behind the star begins to drop. 
As the mass-stripping phase nears completion, the companion
is almost spherically symmetric and close to its terminal velocity.

Column three in Table \ref{kickchart} provides our ballpark estimate
of the kick that we calculate by assuming an inelastic collision in which all
the momentum of the supernova ejecta incident on the
remnant of the companion is directly transferred to the remnant.
That is, using $\Delta \Omega_{REM}$ and $M_{REM}$ as the solid angle subtended by the
remnant and the mass of the remnant, and $M_{SN}$
and $V_{SN}$ as the mass and velocity of the supernova ejecta, respectively,
we estimate the kick as
\begin{math}
v \lesssim \Delta \Omega_{REM} M_{SN} V_{SN} / M_{REM}.
\end{math}
This is clearly an overestimate because not all of the momentum will
be directly transferred to the remnant. The 
momentum transferred to the remnant will depend on the
details of the interaction, such as the changing
geometrical area as mass loss decreases the surface area
of the companion.
Our estimate is directly related to the method of Wheeler \etal (1975),
except that we drop the logarithmic factor because
it was concluded (\cite{fryxell81}; \cite{taam84}) that ablation
did not substantially increase the kick.

\section{Systematic Trends with Binary Separation}
\label{sec:separation}

\newcommand{\frameZA}{\ref{ms1_em5}a~}
\newcommand{\frameZAA}{\ref{ms1_em5}a}
\newcommand{\frameZB}{\ref{ms1_em5}b~}
\newcommand{\frameZBB}{\ref{ms1_em5}b}
\newcommand{\frameZC}{\ref{ms1_em5}c~}
\newcommand{\frameZCC}{\ref{ms1_em5}c}
\newcommand{\frameZD}{\ref{ms1_em5}d~}
\newcommand{\frameZDD}{\ref{ms1_em5}d}

	To study the effect of changing the binary separation, we performed
four test cases with the $1.0$ \Msun main sequence secondary 
in addition to the HCV simulation:
HCVa, HCVb, HCVc, and HCVd. The ratio of binary separation
to stellar radius ($a/R$) is $2.57$, $4.0$, $6.0$, and $12.0$,
respectively, with HCV having a ratio of $3.0$, making it the
second-closest secondary (see Table \ref{sims}). 
Because the incident momentum scales as the inverse square of the
separation, we expect a significant change in the
amount of stripped mass and in the kick. 
From Tables \ref{stripchart} and \ref{kickchart} we see
that over this range of orbital separation the stripped mass
varies more than a factor of $100$, from $0.23$ \Msun to
$0.0018$ \Msun, and the kick varies by a factor of $9$, from
$137$ \kmsec to $15$ \kmsecc.

Figures \ref{ms1_em5} visually demonstrate
the change in the strength of the impact.
In the strongest impact, the 
companion experiences the strongest initial acceleration, 
followed by the strongest deceleration as the pressure rises
in the back of the star. In the weakest impact, the
companion experiences a very weak acceleration, which
is not followed by a deceleration phase. In this case,
the shock propagating through the companion is too weak 
to cause a pressure rise in the back of the star.
Figures \ref{ms1_em5} illustrate the decrease in strength of the impacts
with a series of 2-D images shown at $1000-1300$ seconds after the
supernova explosion.
In the HCV simulation (Figure \frameZAA), the shock has already
passed through the companion. The bow shock is broad, within which we see 
well-developed \kh and \rt instabilities trailing behind the companion
that indicate mass loss is occurring. 
In the HCVb simulation (Figure \frameZBB), the interaction is at a comparable
phase, except that the shock has just converged behind the star and
the trailing instabilities are less developed.
In the HCVc simulation (Figure \frameZCC), the shock is still passing
through the star and the bow shock is noticeably much narrower than in
the stronger impacts.  In comparison with the other cases,
the impact is almost mild in the HCVd simulation (Figure \frameZDD).
Here, the bow shock is very narrow and the shock's propagation through the star is slow.

We use the separation study to compare our
numerical stripped mass and kick with the
analytic formulae of Wheeler \etal (1975).
From Tables \ref{stripchart}, \ref{kickchart}, and \ref{sims} we see
that the stripped mass and kick are sensitive functions of $(a/R)$.
As shown in the tables, the analytic stripped mass usually 
overestimates the amount of mass loss by $5-50 \%$
for separations in the range of $a/R = 3.0-4.0$. For greater
separations, the mass loss is severely overestimated.
For smaller separations, the mass loss is underestimated.
Our analytic estimate of the kick is calculated from
simple conservation of momentum and does not include
a logarithmic factor to represent an additional kick
from ablation. The analytic method consistently 
overestimates the kick from the impact,
by $18 \% - 47 \%$ in the range of our test cases. 
To conclude, we can say that
the analytic estimates of the stripped mass provide a good estimate 
(within $10 \%$) for binaries near Roche lobe overflow  ($a/R \approx 3$),
but for closer and farther separations the 
estimates are substantially in error. Furthermore, the
analytic estimates of the kicks provide a ballpark value for
all separations, but are systematically high. 
The kicks are below the orbital velocity of the secondary
for all orbital separations.

We can use the separation study
to see how the average drag coefficient is likely to 
depend on the strength of the impact.
The efficiency of the transfer of momentum is given by
the ratio of stellar remnant momentum to the incident
supernova momentum and can be equated to an average drag coefficient
(Table \ref{kickchart}).
Typical coefficients lie in the range of $0.2 - 0.5$,
rising systematically as the binary separation 
increases. We note that,
as the binary separation increases, less mass is
stripped from the companion and its geometrical area does
not decrease as much. More of the incident supernova 
momentum can be intercepted by the remnant.
In essence, the average drag rises for impacts in which
less material is stripped.

Despite the dramatic change in stripped mass and kick,
the characteristic velocity of the stripped material
stays between $855$ \kmsec and $664$ \kmsecc, with the
stripped material from the closest binaries having the 
highest characteristic velocities. 
The half-mass velocities are: $855$ \kmsecc, $823$ \kmsecc, 
$778$ \kmsecc, $664$ \kmsecc, and $730$ \kmsec
for the HCVa, HCV, HCVb, HCVc, and HCVd simulations respectively.
The fluctuations in the high-velocity end of the velocity distributions
seem to reflect the density variations in the initial supernova ejecta model, and 
like the characteristic velocity, decrease in strength and velocity as
the binary separation increases. 

We can understand the characteristic velocity by noting
that the companion's shocked material has a
specific internal energy equal to its specific kinetic energy,
which is determined by the speed of the shock. 
When the shock is strong, the envelope is heated enough
that its new speed of sound exceeds the companion's escape velocity
and the envelope is free to evaporate away from the star.
As the shock decelerates, as it will when it climbs the steep density 
gradient of the star, less specific internal energy is deposited
in the envelope and the new speed of sound will be lower.
Eventually, the speed of sound will be lower than the escape velocity
and the heated material will remain bound to the star.
The material that escapes first will be the material shock-heated in
the outer layers where most of the energy is deposited; this material
will have the highest characteristic velocities. 
In time, the underlying layers of the envelope will evaporate off, leaving
the marginally bound material clinging to the stellar surface.
The amount of mass heated to sufficiently high temperatures is
less for greater separations.

The closest companions will create the widest plume of
stripped material. For an explosion in a close binary
with a hydrogen-rich companion, the contaminating
hydrogen can fill a solid angle of $2 \pi$ steradians.
A companion in a more distant binary will leave a 
narrower trail of contaminating hydrogen.
Figure \ref{polar_em5series} shows the solid angle distribution plotted
as a polar plot, with the radius being the magnitude of 
$dM/d\alpha$ and $\alpha$ once again being the angle from the
downstream axis. Clearly, the closer the secondary is to the explosion, 
the wider in solid angle is the stripped material. 
The half-mass angles for each distribution range from
$10^{\circ}$ to $45{^\circ}$.
As shown in Figure \ref{polar_em5series}, most of the contaminating 
hydrogen is exterior to the geometrical shadow.
The outer extent of the contamination is related to the
amount of mass loss, which as can be seen in Figures \ref{ms1_em5},
is greatest for the strongest impacts.

\section{The Subgiant Companion}
\label{sec:subgiant}

Next, we describe the impact of the
blast wave on a subgiant star, our HCVL scenario.
As in the main sequence case,
we neglect any changes to the structure
of the star from the Roche potential or an
optically thick wind from the white dwarf
which would complicate the simulation.
For this simulation 
we use a $330 (\rho) \times 565 (z)$ cylindrical grid with 
the $1.13$ \Msun subgiant secondary
centered on the origin. 
Approximately $200 (\rho) \times100 (z)$ zones are allocated 
to the secondary.
We use the same dimensions relative to
the companion's radius as the main sequence simulation:
$6$ stellar radii in the $\rho$ direction,
$12$ stellar radii in the downstream (negative $z$) direction
and $2$ stellar radii in the upstream (positive $z$) direction.
The same supernova ejecta model (W7) is used and the debris
flows onto the grid in the same way.
We show time frames of this simulation up to
$2.0\times10^{4}$ seconds, which is longer than the 
companion's dynamical time of $3.5\times10^{3}$ seconds
and longer than 
the $2.5\times10^{3}$ seconds that
it takes for the trailing edge of the supernova
to sweep past the back of the companion.

For clarity, this section is subdivided into distinct topics.
In \S \ref{sec:hydrosg} we describe the hydrodynamic stages of the impact
and in \S \ref{sec:stripsg} we describe the stripped mass for
this scenario. 
In \S \ref{sec:distsg}, we discuss the velocity and solid angle distributions
of the stripped companion material and the observational 
implications.
Finally, in \S \ref{sec:kicksg}, we discuss the
kick received by the remnant of the companion.
To avoid repetition, we focus only on the important
differences between the main sequence and subgiant simulations.

\subsection{Hydrodynamics of the Impact on the Subgiant Companion}
\label{sec:hydrosg}

\newcommand{\frameSA}{\ref{sg1}a~}
\newcommand{\frameSAA}{\ref{sg1}a}
\newcommand{\frameSB}{\ref{sg1}b~}
\newcommand{\frameSBB}{\ref{sg1}b}
\newcommand{\frameSC}{\ref{sg1}c~}
\newcommand{\frameSCC}{\ref{sg1}c}
\newcommand{\frameSD}{\ref{sg1}d~}
\newcommand{\frameSDD}{\ref{sg1}d}

The impact of the supernova W7 on the subgiant follows the
same sequence of events as the impact on the main sequence star.
The subgiant subtends almost the same solid angle as the
main sequence star (see Table \ref{sims}), so the incident momentum
is almost the same in both simulations. 
Given this, it is not surprising that the character of the impact
has not changed. However, because of the increase in
physical scale, which goes as the radius of the companion star,
the sequence of events occurs at slightly different times. 
Also, the compact core of the subgiant alters the progress of
the shock through the star, and the larger envelope
increases the timescale required to bring the remnant back into
hydrostatic equilibrium.
The highlights of the impact on the subgiant are shown in
several 2-D images in Figures \ref{sg1} and \ref{sg2}.

Figure \frameSA shows the impact of the supernova ejecta on the
subgiant 343 seconds after the explosion. 
The leading shock front preceding the ejecta has converged
along the downstream axis. The density jumps in the ejecta
that are visible as spherical rings have already distorted the 
bow shock. 
More than $90 \%$ of the supernova's mass and momentum that will
collide with the subgiant is now represented on the grid.
At this early stage, only $0.06$ \Msun of the
secondary's envelope has been stripped and the stripped 
material is still in close proximity to the secondary.

Figure \frameSB illustrates the curvature of the shock front
at $945$ seconds.
As with the main sequence companion, the
shock front driven through the center of the subgiant decelerates
as it nears the subgiant's tightly-bound core. The shock front driven 
into the outer envelope propagates so much faster that it converges
on the downstream axis just after the stellar core is shocked.
As with the main sequence companion, a reflection shock passes back
through the subgiant. 

By Figure \frameSCC, $2044$ seconds after the
explosion, the supernova shell has swept by the
secondary. Shock-heated material can be seen 
streaming from the subgiant within the cavity
shaped by the bow shock. By this time,
$0.15$ \Msun of stellar material has been stripped.
The \rt and \kh instabilities can be seen easily in Figure \ref{sg2}.

At $2\times10^{4}$ seconds (Figure \frameSDD), the companion is
beginning to recover hydrostatic equilibrium. Like the
main sequence companion, the star is surrounded by an extended
envelope of low-density material. A complicated series of
shocks formed during the stellar pulsation phase are
propagating slowly through the outer envelope.
By this time, $0.17$ \Msun has been stripped and the subgiant has
received a kick of $49$ \kmsec from the impact.

The pressure and density of the subgiant fluctuate in
the same way as in the main sequence case. 
The central pressure first rises rapidly as the
shock passes through the stellar core at $600-700$ seconds, then
slowly falls as the star expands.
As in the main sequence impact, the material which is now
only marginally bound begins to collapse back onto
the subgiant and at $\sim 4000$ seconds
a shock is transmitted into the outer envelope.
A pulsation phase follows.

The density and entropy profiles for the pre- and post-impact
subgiant are shown in Figures \ref{Dens1} and \ref{S1}.
The radius of the remnant star, which we define as
the radius that encloses about $90 \%$ of the remnant's mass,
is $\sim 0.9$ \Rsunn, which is a decrease of $48 \%$. 
The remaining $10\%$ of the mass forms a low-density extended envelope 
out to $5.9\times10^{6}$ km ($\sim 8$ \Rsunn).

\subsection{The Stripped Mass of the Subgiant Companion}
\label{sec:stripsg}

We determine the amount of mass stripped in the same way we did for the 
main sequence case and list it in Table \ref{stripchart}.
We include the mass stripped for two subgiant scenarios: HCVL and HCVLa.
The HCVL binary scenario uses a 
$1.13$ \Msun subgiant which was created from a $2.1$ \Msun
subgiant that was artificially stripped of its outer envelope
to coarsely mimic mass transfer prior to the supernova explosion.
The HCVLa binary scenario uses the original $2.1$ \Msun subgiant
with a slightly larger binary separation so that the same 
solid angle is subtended in order to ensure that both secondaries receive the
same incident momentum from the blast.
This scenario was included only as a test case.
At the end of the HCVL simulation, $3\times10^{4}$ seconds
after the supernova explosion ($\sim 5$ sound crossing times),
$0.17$ \Msun of stellar material has become unbound, and
of this material, about $97\%$ has left the grid by the last
timestep. 
The mass-stripping phase lasts roughly $3.7\times10^{3}$ seconds,
long enough for $96 \%$ for the
mass that will be stripped by the end of the simulation to
be stripped,
and much longer than the $2.5\times10^{3}$ seconds required for 
passage of the supernova blast around the secondary.
In comparison, the original $2.1$ \Msun subgiant in
the HCVLa simulation loses $0.25$ \Msunn.

Table \ref{stripchart} compares the numerical stripped mass with
an analytic estimate. 
The analytic calculation, which underestimates the stripped mass
for both the HCVL and HCVLa simulations, is nonetheless 
within $22\%$ of the numerical result.
We ran the HCVLa simulation as a test case to see if the stripped
mass scales directly with the companion's mass for the same
subtended solid angle. We find that the mass loss
scales almost directly with the mass of the companion.
That is, the $2.1$ \Msun subgiant loses $0.25$ \Msunn, which is
$12\%$ of its mass. In comparison, the $1.13$ \Msun subgiant
loses $0.17$ \Msunn, which is $15\%$ of its mass.
            
\subsection{Distribution in Velocity and Angle of the Material Stripped from the
            Subgiant Companion}
\label{sec:distsg}

Our key result for the subgiant simulation is the velocity
distribution of the stripped hydrogen, shown in
Figure \ref{dmdv_bw2}b with the SN W7 velocity distribution. 
The velocity profile of the material stripped from the subgiant
shows many of the same features seen in the 
main sequence star simulation.
The half-mass point for the subgiant is just
slightly faster: $890$ \kmsec compared
with $820$ \kmsec for the main sequence case.
Both characteristic velocities are much less than the half-mass
velocity of the supernova ejecta of $7836$ \kmsecc.
Figure \ref{dmdv_bw2}b marks the characteristic
velocities with a vertical line and an arrow.
As in the main sequence case, the bulk of the stripped
material, predominately hydrogen and helium, 
is embedded within the iron layer of the supernova ejecta.
Once again, a high-velocity tail exists which implies that
a trace of companion hydrogen is swept up into the oxygen and
silicon layers of the supernova ejecta.

The differences between the main sequence and
subgiant can be seen in Figures \ref{dmdv_bw2},
which shows the asymptotic velocity profile.
As discussed in \S \ref{sec:distms}, we subtract
in quadrature an estimate for the local escape velocity that
was used to distinguish between the stripped and the unstripped
material. This approximates the velocity profile at infinity,
when the stripped material is free of the secondary's
gravitational potential. 
The velocity profile of the mass stripped from the subgiant
companion shows a subtle shift to higher
velocities, and the high-velocity tail shows a 
single peak near $3\times10^{3}$ \kmsecc.
The velocity profile of mass stripped from the main sequence companion shows
three peaks, which we identify with the three composition transitions
in the W7 model; the other two peaks in the mass stripped from the
subgiant are not clearly distinguishable.

Just as the velocity profile of the subgiant's stripped material
is very similar to the velocity profile of the main sequence star's
stripped material, the spatial distribution of the subgiant's
stripped material shows a strong
similarity to that in the main sequence case.
Figure \ref{dmds_ts85_ts88} illustrates the distribution in solid angle
(dM/d$\Omega$) of the stripped material in the HCVL and HCV simulations, 
expressed in terms of the angle, $\alpha$, from the downstream axis.
The half-mass point of the HCVL distribution is $\sim 49^\circ$,
just slightly more extended in angle than in the main sequence case,
which has a half-mass angle of $43^{\circ}$.
In Figure \ref{polar_ts85_ts88},
the polar angle, $\alpha$, once again represents the angle from the downstream 
axis and the radial coordinate represents the magnitude of $dM/d\alpha$.
The material stripped from the subgiant and the main sequence companions
fills almost the same cone, with the subgiant filling in a slightly 
broader one. Once again,
the majority of the stripped mass is exterior to the geometrical shadow,
the shadow that the secondary would cast from the site of the
explosion of the primary. 
The collision of the supernova debris with the subgiant secondary will create
a hole in the debris structure of $\sim 32^{\circ}$ in the high-velocity
ejecta and of $\sim 40^{\circ}$ in the low-velocity ejecta.
The asymmetry of the supernova ejecta will be slightly larger
at the highest velocities for the subgiant model than the main sequence model.
The asymmetry at low velocities will be almost the same for both models.
As in the main sequence model, the asymmetry of the supernova ejecta
is a direct indication of the presence of a companion star.

The hydrogen mass fraction is shown 
as a function of supernova mass and velocity 
in Figures \ref{Mts88} and \ref{Vts88}.
The hydrogen mass fraction falls abruptly 
as the velocity increases. 
At $10^{4}$ \kmsecc, which is near the oxygen and
silicon layers, X(H)$\sim 3\times10^{-3}$. 
Of course, we know from the numerical simulation that the
ejected material contaminates a wide solid angle behind
the secondary. For the HCVL simulation, $90\%$ of the
stripped hydrogen lies within $72.5^{\circ}$ of the
downstream axis,
which corresponds to a fractional solid angle of $35\%$.
Figures \ref{Mts88} and \ref{Vts88} show the expected 
mass fraction within the contaminated solid angle.
As expected, the mass fraction rises dramatically.
Of course, if the subgiant were of higher mass, 
then the hydrogen mass fraction 
in the supernova ejecta will rise accordingly.
A higher mass secondary may be physically reasonable
because our subgiant was initially $2.1$ \Msun before
its envelope was modified to mimic mass transfer prior
to explosion.
  
\subsection{The Kick Received by the Subgiant Companion}
\label{sec:kicksg}

The net kicks for the HCVL and HCVLa subgiant scenarios are
listed in Table \ref{kickchart} along with the kicks for
the main sequence companions.
For the HCVL simulation, we see the same qualitative behavior as in 
the main sequence simulations. The secondary receives a strong kick in
the first $700$ seconds after the explosion. A brief period of deceleration
occurs until $\sim 1700$ seconds, and the secondary is left with a 
terminal velocity of $49$ \kmsec by $4000$ seconds.
The numerical velocity is only $70 \%$ of the analytic value of
$70$ \kmsecc.

\section{The Red Giant Companion}
\label{sec:redgiant}

In this section, we describe the impact of a
blast wave with a red giant star, our HALGOLa and SYMB scenarios.
These scenarios differ in the explosion model and binary separation
(see Table \ref{sims}). The SYMB scenario has a slightly larger
binary separation because the mass transfer is by wind accretion,
and its white dwarf primary explodes as a \subCh Type Ia. The HALGOLa
scenario has a binary separation consistent with Roche lobe overflow,
and its white dwarf explodes as a \Ch Type Ia. 
We use the same $0.98$ \Msun red giant model for both simulations.	
As in the main sequence and subgiant cases, we neglect any complicating 
changes to the structure of the envelope from the Roche potential.

We change to spherical coordinates for all the red giant
simulations.
The grid extends $12$ stellar radii and $\pi$ radians.
We use $585 (r)\times300 (\theta)$ zones for the HALGOLa
simulations and $653(r)\times300 (\theta)$ zones
for the SYMB simulation, with $242 (r)\times300 (\theta)$ and
$247 (r)\times300 (\theta)$ zones reserved for the red giant's envelope.
The slight difference in the number of zones results from changing 
the grid to accommodate the different
density and velocity profiles of the SN W7 (\cite{nomoto84}) 
and SN Hedt (\cite{woosley94}) ejecta models. 
We run the SYMB simulation for $8.0\times10^{6}$
seconds and the HALGOLa simulation for $6.0\times10^{6}$
seconds, which is only slightly more than one dynamical
time, but still plenty of time for the stripped material
to flow off the hydrodynamic grid.
Because the loosely-bound envelope of the red giant
is almost entirely stripped by the impact, there is no
need to extend the calculation any further.
We will concentrate on the SYMB simulation in 
this discussion, but will point out differences
with the HALGOLa scenario when appropriate.

For clarity, this section is subdivided into distinct topics.
Section \ref{sec:hydrorg} describes the character of the impact
and \S \ref{sec:striprg} discusses the stripped mass
for this scenario. 
In \S \ref{sec:distrg}, the velocity and solid angle distributions
of the stripped companion material and the observational 
implications are discussed.
We focus only on the important differences with
the main sequence and subgiant simulations discussed earlier.

\subsection{Hydrodynamics of the Impact on the Red Giant Companion}
\label{sec:hydrorg}

Because of the lower binding energy of the
envelope of the red giant (Figure \ref{be}),
the impact of the supernova on the red giant companion
is much more dramatic than the impact on the main sequence
and subgiant companions.
Unlike the main sequence and
subgiant secondaries, the incident kinetic energy ($\sim10^{49}$ ergs)
is over $1000$ times the envelope binding energy ($\sim10^{46}$ ergs).
The catastrophic effect of the blast on the weakly-bound envelope
was pointed out by a number of authors
(cf.~\cite{wheeler75}; \cite{chugai86}; \cite{applegate89})
and demonstrated numerically by Livne \etal (1992).
Our simulations confirm the basic result of
Livne \etal (1992): the envelope of the red giant is stripped
and the velocity of the stripped material is less
than $10^{3}$ \kmsecc. The half-mass point of our velocity distribution is 
$600$ \kmsecc.
Like Livne \etal (1992),
we select a binary scenario
with the red giant close enough to the white dwarf 
to be in Roche lobe overflow, our HALGOLa, and a scenario with a 
slightly larger binary separation, our SYMB.
Unlike Livne \etal (1992), instead of a parameterized debris model,
we use supernova Type Ia
W7 for the blast for the HALGOLa and Type Ia
\subCh Hedt for the SYMB (Table \ref{ejectamodels}).
Figures \ref{rg1} and \ref{rg2} show highlights
of the destruction of the red giant's envelope in a series of
2-D images from the SYMB simulation.
However, Figures \ref{rg1} focus on the propagation of the
shock through the envelope, while Figures \ref{rg2} show the entire grid.

\newcommand{\frameRA}{\ref{rg1}a~}
\newcommand{\frameRAA}{\ref{rg1}a}
\newcommand{\frameRB}{\ref{rg1}b~}
\newcommand{\frameRBB}{\ref{rg1}b}
\newcommand{\frameRC}{\ref{rg1}c~}
\newcommand{\frameRCC}{\ref{rg1}c}
\newcommand{\frameRD}{\ref{rg1}d~}
\newcommand{\frameRDD}{\ref{rg1}d}

\newcommand{\frameTA}{\ref{rg2}a~}
\newcommand{\frameTAA}{\ref{rg2}a}
\newcommand{\frameTB}{\ref{rg2}b~}
\newcommand{\frameTBB}{\ref{rg2}b}
\newcommand{\frameTC}{\ref{rg2}c~}
\newcommand{\frameTCC}{\ref{rg2}c}
\newcommand{\frameTD}{\ref{rg2}d~}
\newcommand{\frameTDD}{\ref{rg2}d}

To start the simulation,
the red giant is centered at the origin, and the density and
velocity profiles of the exploding white dwarf, scaled to $1.73$ hours
after the explosion, are positioned along the upstream axis.
The impact begins at $\sim 3.0$ hours after the explosion.
Figures \frameRA and \frameTA show the interaction 
$5.67$ hours later, at $8.68$ hours after the explosion.
The initial impact has driven a shock into the companion.
The shock front in the ``circumstellar'' medium
joins smoothly with the shock propagating through the 
red giant's envelope. The supernova material has not yet converged 
on the downstream axis.

By $14.23$ hours, the bow shock has formed
and the supernova ejecta have converged on the downstream axis
right behind the star (Figures \frameRB and \frameTBB).
The shock passing through the companion is extremely curved.
\kh and \rt instabilities are clearly visible in the shock-heated
material.
The faint line of low-density material that is along the
upstream axis and just outside the bow shock
is a numerical artifact caused by the reflecting boundary. 

After $2.10$ days (Figures \frameRC and \frameTCC), 
the shock moving through the envelope converges in the back 
of the star. The entire star has been shocked, and 
almost the entire envelope has now been stripped down 
to the degenerate core.
The slight distortion of the bow shock is due to the 
numerical artifact along the upstream axis. 
A reflection shock will soon be created that will pass back 
through the star.

By $5.86$ days, pressure behind the bow shock
has pushed it back towards the site of the explosion
(Figure \frameRD and \frameTDD).
Although the velocity of the stripped material near the core
is almost spherical, the bulk of the stripped material
is moving downstream. In this simulation the
bow shock is showing distortions as the density drops
in the post-shock region.
At this point, we regrid to increase the timestep
and continue the simulation for an additional $\sim 86.7$ days,
or $92.6$ days after the explosion, which is time enough for 
$\sim 85 \%$ of the stripped material to flow off the grid.
Only a trace amount of material, at most $4\%$ of the 
original envelope ($\sim 0.02$ \Msunn), is left clinging to 
the degenerate core.

Unlike the main sequence and subgiant companions, which are
done in cylindrical coordinates with the supernova debris
added to the grid by the boundary conditions, 
the red giant simulations require the density and velocity
profiles to be placed directly on the grid, off-center,
as part of the initial conditions. The supersonic flow must 
propagate across radial lines and maintain its spherical
structure. Figures \ref{rg2} show that the spherical structure
of the debris is maintained. But does placing the spherical
explosion off-center have any consequences for the amount of
stripped mass or the velocity distribution? 
To test this and to address a concern that
the numerical artifact along the axis might have
unexpected consequences, we inverted the position of the 
supernova and the secondary for the HALGOLa simulation. 
Although there are small differences between
the two simulations, 
in both cases over $0.54$ \Msun of the red giant's 
envelope is ejected by the blast. For the original HALGOLa
simulation the half-mass point of the velocity distribution of the
stripped material is $593$ \kmsecc, and for
the inverted simulation the half-mass velocity is $584$ \kmsecc.

After the red giant's envelope is ejected, the degenerate
core, now a single low-mass He pre-white dwarf, 
is left surrounded by a hot, entropized atmosphere.
Although the blast strips most of the envelope, a small
residual fraction (a few percent) may form an extended, hydrogen-rich
envelope around the star.
Although about $\sim75\%$ of this envelope is within
$4.4\times10^{7}$ km and has typical temperatures
of $10^{4} - 10^{5}$ K, the remaining $\sim 25\%$ 
is cooling adiabatically, and extends out as far as $2.4\times10^{8}$ km.
Typical entropies in the envelope are $35-40$ in units of k$_{B}^{-1}$ baryon$^{-1}$.
In time, this extended envelope will settle onto the
star, adding $\sim 0.02$ \Msun of hydrogen-rich material to the
hydrogen burning layer. We speculate on the fate of the
low-mass He pre-white dwarf in \S \ref{sec:postimpact}.

\subsection{The Stripped Mass of the Red Giant Companion}
\label{sec:striprg}

We determine the mass lost in the same way
we did for the main sequence and the subgiant cases.
The stripped mass for the red giant simulations is shown in
Table \ref{stripchart} with the results from the other simulations.
Like Livne \etal (1992),
we find that almost the entire envelope is
removed by the impact, for all the red giant binary scenarios.
For the HALGOLa simulation,
the mass-stripping phase lasts roughly $10^{5}$ seconds (or $1.2$ days).
After $6.0 \times10^{6}$ seconds (or $69.5$ days),
$0.54$ \Msun ($98 \%$ of the stellar envelope) of stellar material has 
become unbound, and of this material, about $85\%$ has left the grid.
The stripped mass for the SYMB simulation is similar.
The mass-stripping phase for the SYMB simulation  lasts roughly $1.3\times10^{5}$ seconds
(or 1.5 days).
After $8.0 \times10^{6}$ seconds (or $92.6$ days),
$0.53$ \Msun ($96 \%$ of the stellar envelope) of stellar material has 
become unbound, and of this material about $85\%$ has left the grid. 

\subsection{Distribution in Velocity and Angle of the Material Stripped from the
            Red Giant Companion}
\label{sec:distrg}

Figures \ref{dmdv_bw1} show the
velocity distribution of the stripped envelopes; both are shown
with the original velocity distribution of the supernova ejecta,
W7 for the HALGOLa simulation and Hedt for the SYMB simulation.
The velocity distributions from the red giant simulations, which have
the same basic shape as the distributions from the
main sequence and subgiant simulations,
have systematically shifted to lower velocities. 
The half-mass velocity of HALGOLa is $593$ \kmsec compared with
$421$ \kmsec for the SYMB case.

Because the Hedt debris profile has only $78\%$ of the momentum
of the W7 profile, we expect the SYMB scenario to have 
a lower characteristic velocity. In fact, the SYMB stripped material 
has a half-mass velocity of $71\%$ of the HALGOLa case.
However, the SYMB scenario has a slightly greater separation.
By repeating the SYMB simulation with the same binary separation as
the HALGOLa simulation, we find that 
the solid angle distribution remains almost unchanged, but
the new velocity profile is almost, but not quite, identical to
the HALGOLa profile.  As in the HALGOLa distribution, the new 
half-mass velocity is $593$ \kmsecc.
The only distinct difference between the velocity profiles is 
the slight bump in the high-velocity tail (near
$\sim 3.0\times10^{3}$ \kmsec in Figures \ref{dmdv_bw1}),
which is visible in the HALGOLa case and not visible in
either the old or the new SYMB case.
We conclude from this that the slight increase in the high-velocity
tail reflects the different supernova ejecta models used and that
the shift to lower velocities in the SYMB case reflects the
slightly greater binary separation.

The solid angle distribution for both of the red giant scenarios
is shown in Figure \ref{dmds_t117_t120_sn} and as a polar plot
in Figure \ref{polar_t117_t120}.
Although the $dM/d\Omega$ profiles
appear to be rising along the downstream axis, the amount of
mass ejected in that direction is actually less than would be expected
if the envelope had been uniformly stripped. In fact, 
the half-mass angles are $61^\circ$ and $66^\circ$, with that for
the SYMB scenario being slightly more extended.
As in the main sequence and subgiant cases, the majority of
the stripped mass is exterior to the solid angle
enclosed by the geometrical shadow of the secondary as seen from
the site of the explosion. In addition, there is now 
a substantial amount of mass that has been ejected 
past the $90^\circ$ point.

The impact of the supernova ejecta onto the secondary creates a
hole in the supernova debris (Figure \ref{dmds_t117_t120_sn}).
For the HALGOLa model, the supernova debris has a hole
of $\sim 34^{\circ}$ in the outer ejecta, which widens
to $\sim40^{\circ}$ in the inner ejecta.
This corresponds to $9\%-12\%$ of the surface of the ejecta.
The SYMB model has a slightly smaller hole
of $\sim 32^{\circ}$ in the outer ejecta, which widens
to $\sim37^{\circ}$ in the inner ejecta.
This corresponds to $8\%-12\%$ of the surface of the ejecta.
The HALGOL model clearly creates more asymmetry in the
supernova ejecta at the highest velocities than the SYMB model or
the main sequence and subgiant models. 

Figures \ref{Mt117} and \ref{Vt117} illustrate X(H), the
hydrogen mass fraction in the contaminated 
supernova ejecta (assuming uniform angular mixing in the
unperturbed supernova model) for the HALGOLa scenario.
X(H) is generally higher in the red giant simulations than in the
main sequence or subgiant simulations simply because of the
larger mass ejected. However, 
because most of the hydrogen stripped from the red giant
is at lower velocities, for the red giant case 
the hydrogen mass fraction at high velocities 
is slightly lower than in the main sequence and
subgiant cases in that X(H) $\sim 2\times10^{-3}$ at velocities
of $10^{4}$ \kmsecc, compared with X(H) $\sim 3\times10^{-3}$ for
the main sequence and subgiant cases.
If we constrain the mixing to just the contaminated solid angle
behind the secondary, X(H) does
rise, but not dramatically, because the
contaminated solid angle is now so much larger
that this correction becomes small.
For the HALGOLa simulation, $90\%$ of the stripped hydrogen 
lies within $115.4^{\circ}$ of the
downstream axis which corresponds to a fractional solid angle of $71\%$.

\section{Upper Limit on  High-Velocity Hydrogen}
\label{sec:spectrum}

Ideally, we would like a firm upper limit on the amount of
hydrogen that can be mixed with the supernova debris at high and low
velocities without being detected in the spectrum. The low-velocity
upper limit requires a non-LTE radiative transfer study (\cite{pinto99}).
For the high-velocity upper limit we can use upper limits
from Type Ia observations near maximum light. Although most of the
stripped material is ejected at low velocities, for all the
binary scenarios we have considered, there is a small high-velocity tail,
as shown in Figure \ref{highv_bi}. 
In Table \ref{highvchart}, selected points are tabulated.
We present the velocity distribution as the total amount of hydrogen 
exterior to each velocity.\footnote{
In \S \ref{sec:numericalmethods}, we discuss the effect of
resolution on the high-velocity profile.
In the HCV simulation there is $15\%$ more mass above
$3\times10^{3}$ \kmsec and $26\%$ less mass above 
$1.5\times10^{4}$ \kmsec than in the high-resolution simulation.
Hence, at the higher velocities we conclude that changes in mass greater 
than $\sim30\%$ are meaningful.}

Above $3.0\times10^{3}$ \kmsecc, the left-hand edge of Figure \ref{highv_bi},
all four scenarios have less than $\sim0.014$ \Msun of high-velocity hydrogen.
The HCVL simulation has $0.014$ \Msunn, followed by the
HALGOLa with $0.012$ \Msunn, the HCV with $0.010$ \Msunn, and lastly, the SYMB simulation
with only $0.004$ \Msunn. 
By $10^{4}$ \kmsecc, the amount of high-velocity hydrogen for all four scenarios has
dropped by almost a factor of 10. 
The HCVL, HCV, and the HALGOLa scenarios are very similar up to
this velocity. Because of the greater binary separation and smaller impact 
momentum and energy, the SYMB scenario has less than half the hydrogen of the
other scenarios. When the SYMB scenario has the same binary separation
as the HALGOLa scenario, it too has a velocity distribution like
the others in this range ($3\times10^{3}-10^{4}$ \kmsecc).
It would be difficult to discriminate between the main sequence, the
subgiant, and the HALGOLa red giant scenarios in this velocity range. 
However, we can easily discriminate between the companions in 
Roche lobe overflow and a model in which the red giant that is too 
distant to be in Roche lobe overflow, our SYMB scenario.

For $1.5\times10^{4}$ \kmsecc, the HALGOLa and the SYMB scenarios have
the most high-velocity hydrogen, the HALGOLa with $3.3\times10^{-4}$ \Msun
and the SYMB with $3.0\times10^{-4}$ \Msunn. 
In Figure \ref{highv_bi}, the hydrogen distribution for the
SYMB and HALGOLa scenarios converge at higher velocities, implying that
the HALGOLa and the SYMB scenarios have almost
the same amount of hydrogen at the highest velocities.
We attribute this convergence to the momentum profile of 
Hedt (\subCh model), which has more overall momentum at velocities 
greater than $1.5\times10^{4}$ \kmsec than SN W7 (\Ch model)
(Figure \ref{SN}). 
We note that the amount of high-velocity hydrogen
above $1.5\times10^{4}$ \kmsec could be a direct indicator of the
momentum profile of the supernova ejecta at the highest velocities.

Above $1.5\times10^{4}$ \kmsecc, the HCVL and HCV scenarios have much less
high-velocity hydrogen than in the red giant scenarios.
The HCVL scenario has only $1.7\times10^{-4}$ \Msunn, which is only $53\%$ of
the high-velocity hydrogen of HALGOLa. The HCV has even less, only
$1.3\times10^{-4}$ \Msunn. 
We note that even if the red giant companions
have more low-velocity hydrogen, they have more hydrogen
above $1.5\times10^{4}$ \kmsec than the main sequence or subgiant models.
Because the HCVL has more high-velocity hydrogen than the HCV scenario, in principle
we can discriminate between the subgiant and main sequence companion
at $1.5\times10^{4}$ \kmsecc.

The upper limit of H/Si $\le 2.0\times10^{-6}$ (relative to solar) from SN 1990M,
which, assuming perfect mixing of the silicon in an ejecta profile
like W7, corresponds to a total hydrogen mass of
$\sim 3\times10^{-4}$ \Msun (\cite{dellavalle96}) and is indicated by a
horizontal line in Figure \ref{highv_bi}.
Because Della Valle \etal (1996) estimated
the error in the upper limit to be a factor of $2-3$,
we also include horizontal lines to indicate this range.
The vertical line indicates the velocity in SNIa W7 that corresponds
to the outer $0.04$ \Msun of the ejecta, which they estimated
to be the mass in the photosphere at the time of the
observations.
If $3\times10^{-4}$ \Msun is a reasonable upper limit to the
total amount of hydrogen, and if it is mixed only within
the observable photosphere, then we can compare our mass
profiles with this upper limit at the edge of the observable
photosphere. We see from the figure that none of the binary scenarios 
we have considered exceeds this upper limit.
However, if the upper limit were reduced by only a factor of 3,
the red giant and subgiant models would clearly exceed this limit,
while the main sequence model is marginal.
Because the supernova photosphere is non-LTE and the upper limit of
Della Valle \etal (1996) is based on an LTE atmosphere,
the hydrogen abundance in the supernova could be much higher than their
upper limit indicates.

\section{The Future of the Companion Star}
\label{sec:postimpact}

The post-impact evolution of the main sequence and subgiant secondaries
is a separate calculation requiring a 2-D stellar evolution
code that is capable of handling initial models
that are slightly asymmetric in density and temperature and
completely out of thermal equilibrium.
For example, although the main sequence star is recovering 
hydrostatic equilibrium by the end of the simulation, 
the density in the outer envelope of the  star still varies by 
as much as a factor of $2$ in different directions.
The temperature and entropy profiles are also highly asymmetric,
reflecting the shock history of the impact. 
Without attempting to follow the secondary's post-impact evolution 
in detail, we can only speculate about the secondary's future evolution.

Immediately after the impact, the main sequence star
is puffed up, much like a pre-main sequence star.
Although the nuclear energy generation in the core has been extinguished, 
the luminosity of the residue will be dramatically brighter 
for this extended envelope that is out of thermal equilbrium.
Chaboyer (1998) estimated 
that with the asymmetrical temperature distribution
the luminosity could vary from $500$ \Lsun to $5000$ \Lsun after the impact,
with a \kh timescale of $1400$ years to $11000$ years.
After thermal equilibrium is reestablished, the remnant will
return to the main sequence along a \kh track and then will continue its
evolution at a rate prescribed by its new mass. Without mixing in
the stellar core to refresh the hydrogen supply, the born-again star will 
appear near the middle of its main sequence lifetime. Using the main sequence relations 
R $\propto$ M$^{2/3}$ and L $\propto$ M$^{5}$,
the $1.0$ \Msun main sequence secondary will eventually return to the main 
sequence with a mass of $0.85$ \Msunn  (having lost $0.15$ \Msun by the impact),
with a slightly smaller radius of $0.9$ \Rsunn,
and a luminosity near $0.4$ \Lsunn.

The subgiant puffs up after the impact and must contract to 
re-ignite hydrogen shell-burning, which will have been extinguished
during the subgiant's expansion. As in the main sequence case, the luminosity will be 
extremely high during this phase. After thermal equilibrium is established and
burning is reignited,
the star will be back on a post-main sequence track with a slightly
lower luminosity.

After the red giant's envelope is stripped by the 
impact of the supernova, the degenerate
core, now a single low-mass He pre-white dwarf, 
is left surrounded by a hot, entropized hydrogen/helium atmosphere.
The structure of the core is only minimally affected
by the impact because the converging shock was
weakened by the extremely strong pressure and 
density gradients in the interior of the red giant.
Although the blast strips most of the envelope, a small
residual fraction ($\sim 0.02$ \Msunn) may form an extended, 
hydrogen-rich envelope.
Although the impact may temporarily
disrupt the hydrogen-burning layer, it is possible that
the star may recommence shell burning as the envelope
settles back down into a thin layer.
As the degenerate core and hydrogen-rich envelope contracts,
the star will evolve away from the red giant branch,
along a track of constant luminosity ($\sim10^{3}$ \Lsunn) 
and rising effective temperature on a timescale of $10^{5}-10^{6}$ years
for a $0.42$ \Msun He core (\cite{iben86}; \cite{iben93}; \cite{iben93b}).
After the effective temperature rises to greater than $\sim 3\times10^{4}$K,
the star may appear as an underluminous main sequence O or B star. 
At the end of the contraction phase, the luminosity begins to drop.
When the luminosity drops to $\sim10$ \Lsunn, it will appear as
a subluminous hot star; an sdO star if the temperature exceeds $3.5\times10^{4}$K
or an sdB star if it does not (\cite{iben93}; \cite{green99}). 
The star will then continue to cool along a standard He white dwarf cooling 
track (\cite{benvenuto99}). This scenario could be a possible pathway 
for the formation of a subset of single, low-mass He white dwarfs.

\section{Summary and Conclusions}
\label{sec:conclusions}

In this paper, we have presented numerous hydrodynamic simulations
of the impact of a Type Ia supernova explosion with hydrogen-rich main sequence, 
subgiant, and red giant companions. The binary parameters were chosen
to represent several classes of single-degenerate Type Ia progenitor 
models that have been suggested in the literature. All of the simulations 
involved low-mass ($1.0-2.0$ \Msunn) companions that are close enough, 
or almost close enough, to be in Roche lobe overflow when the white dwarf 
primary explodes.

We described the supernova-secondary interaction
for a main sequence, subgiant, and red giant companions with 2-D illustrations
from the simulations. The collision follows the same basic pattern for the 
main sequence and subgiant because of their similar structure.
The initial impact of the supernova shell with the surface 
of the main sequence star drives a shockwave into the stellar envelope.
A reverse shock propagates back into the ejecta. A contact discontinuity
marks the interface between the supernova ejecta and the shocked
stellar material. The shock propagating through the stellar
envelope decelerates as it runs up the steep stellar density gradient.
The shock front, highly curved
because of the density gradients, converges in the back of the star. 
A bow shock develops in front of the companion star.
After most of the supernova debris has passed by, the outer layers
of the stellar envelope are ejected because they have been shock-heated to 
such an extent that the new speed of sound exceeds the companion's
escape velocity. \kh and \rt instabilities are very visible at this stage.
The outer shocked material evaporates away from the
star, embedded within the inner layer of the supernova ejecta.
The stellar core expands and cools in response to the compression 
by the shock's passage through the core. 
The star has an extended, asymmetrical envelope and
begins to pulsate as it settles back into
hydrostatic equilibrium. 

The subgiant follows a similar sequence,
except that its more compact core alters the progress
of the shock through the star and its larger envelope
increases the timescale required to bring the remnant back into
hydrostatic equilibrium. Because of its weakly-bound envelope,
the impact on the red giant is quite different.
The shock propagates through its
more tenuous envelope and converges on its rearward side. All
but a small fraction of the entire envelope is ejected, leaving
a He pre-white dwarf surrounded by a halo of high-entropy hydrogen and
helium.

We can summarize our results by stating that
as a result of the impact of the supernova shell
the main sequence companion loses $0.15$ \Msun and the subgiant companion loses
$0.17$ \Msunn.
In contrast, the red giant companions lose almost their entire envelopes,
at least $0.53$ \Msunn.
The main sequence companion receives a kick of $86$ \kmsecc, the
subgiant $49$ \kmsecc.  In all cases, the kick received by the remnant 
is smaller than the original orbital velocity.
Because it is too small to intercept more than a negligible amount of momentum,
the red giant core will not receive an appreciable kick.

The characteristic velocity of the stripped hydrogen 
(half-mass point in the distribution)
is less than $10^{3}$ \kmsec for all the scenarios.
For the red giant case, the characteristic velocity is in the
range $420 - 590$ \kmsecc, depending on the scenario.
For the main sequence and subgiant cases, 
the characteristic velocities are $820$ \kmsec and $890$ \kmsecc, respectively.
With such low velocities, the bulk of the stripped hydrogen and helium is embedded
within the low-velocity iron of the supernova ejecta.
The region behind the stellar remnant that is contaminated with
stripped hydrogen is always much larger than the geometrical
shadow, the solid angle subtended by the companion as seen 
from the site of the explosion.
For the main sequence companions the stripped hydrogen extends
as far as $66^{\circ}$ away from the downstream axis and
for the subgiant case as far as $72^{\circ}$.
The hydrogen from the red giant contaminates a much larger solid angle,
as far as $115^{\circ}$ from the downstream axis.
These angles correspond to fractions of the sky of $29\%$, $35\%$, and $71\%$.
This is one distinction between the main sequence, the subgiant, and the
red giant scenarios. The stripped hydrogen and helium is distributed
over a much wider solid angle for the red giant scenarios.
Despite the change in distribution of the stripped material,
the hole in the supernova debris (of angular size $31^{\circ}-34^{\circ}$) 
caused by the impact with the secondary is similar for all the scenarios.

If we define the efficiency of momentum transfer as the ratio of the momentum 
transferred to the stellar remnant to the incident momentum, then typical values
for the main sequence simulations lie in the range of $0.2-0.5$,
with the highest efficiency belonging to the secondary with the greatest 
binary separation.  
The characteristic velocity of the stripped material stays within
$660-860$ \kmsecc, with the material from the closest binary companions having
the highest velocities.
The closest binaries will also create the widest plume of stripped material.

One motivation for this project is the possibility of using the
velocity distribution of the stripped hydrogen and helium to discriminate
between Type Ia progenitor models. However, to make 
any definitive predictions requires non-LTE radiative 
transfer calculations using the low-velocity distribution of the stripped 
material to determine the effect of hydrogen/helium contamination on the late-time 
supernova spectrum.
Such calculations can determine the hydrogen and helium line strengths
and ratios of narrow emission lines that will emerge months after the impact, 
as the photosphere recedes to reveal the low-velocity supernova ejecta where 
the bulk of the stripped material is located.
Unfortunately, H$_{\alpha}$ is blended with
numerous Fe and Co lines, especially [\ion{Co}{3}] at $6578$ \AA, and will be
difficult to identify.
Because of atmospheric water absorption, a search for P$_{\alpha}$ ($1.87 \mu m$) can not
easily be conducted from the ground. Searches for P$_{\beta}$ ($1.28 \mu m$) 
can be conducted from the ground, but
telluric absorption and a P$_{\beta}$ emission line from the Earth's atmosphere 
require careful subtraction. Also, P$_{\beta}$ is likely to be blended with a 
broad supernova Fe line at $1.26 \mu m$.
Other lines to look for are He lines at $5876$ \AA, $1.083 \mu m$, and
$2.05 \mu m$ that will be distributed in velocity and angle like the hydrogen,
but note that high-velocity helium is also present in \subCh explosion models.
With a complete study of the emergence of the narrow hydrogen and helium lines
and the competing Fe and Co lines in the late-time supernova spectrum,
an optimum time may be found when the hydrogen may be 
detected and used to distinguish one Type Ia supernova scenario from
another.

Although most of the stripped material is ejected at low velocities, all
the numerical simulations yield a small high-velocity tail.
Above $3.0\times10^{3}$ \kmsecc,
all four scenarios have no more than $\sim0.014$ \Msun of high-velocity hydrogen.
The HCVL simulation has $0.014$ \Msunn, followed by the
HALGOLa with $0.012$ \Msunn, the HCV with $0.010$ \Msunn, and lastly, the SYMB simulation
with only $0.004$ \Msunn. 
Above $10^{4}$ \kmsecc, the amount of high-velocity hydrogen has dropped by
a factor of $10$ for all four scenarios.
In this velocity range ($3.0\times10^{3}-10^{4}$ \kmsecc),
we can easily discriminate between the companions in 
Roche lobe overflow and a model in which the red giant that is too 
distant to be in Roche lobe overflow, our SYMB scenario.
Above $1.5\times10^{4}$ \kmsecc, the HALGOLa and the SYMB scenario have
the most high-velocity hydrogen: the HALGOLa with $3.3\times10^{-4}$ \Msun
and the SYMB with $3.0\times10^{-4}$ \Msunn. 
We note that even if the red giant companions
have more low-velocity hydrogen, they have substantially more hydrogen
above $1.5\times10^{4}$ \kmsec than the main sequence or subgiant models.
Because the HCVL has more high-velocity hydrogen than the HCV scenario,
we can discriminate between the subgiant and main sequence companions
at $1.5\times10^{4}$ \kmsecc.

The Della Valle \etal (1996) observational upper limit from
SN 1990M corresponds to a total hydrogen mass of
$\sim 3\times10^{-4}$ \Msunn.
If this hydrogen is mixed only within
the observable photosphere, then the main sequence, subgiant, 
and the red giant models are just under this upper limit.
However, if the upper limit were reduced by only a factor of 3,
the red giant and subgiant models would clearly exceed this limit,
while the main sequence model is marginal.
With this interpretation, the high-velocity tail of SN 1990M can not yet
be used to exclude any progenitor models with hydrogen-rich companions.

The impact of the supernova ejecta on the secondary star creates
an oddly shaped, small gap in the supernova debris. 
The highest velocity debris flows around the secondary, 
creating a narrow gap of angle $\sim31^{\circ}-34^{\circ}$. 
The lower velocity supernova material flows through the bow shock, 
creating a wider gap in the inner ejecta of angle $\sim40^{\circ}$. 
Therefore, the hole in the supernova
is narrow at the outside of the shell, broad at the inside of the
shell, and filled with high-velocity stripped material from the companion.
As discussed above, most of the stripped material fills a wide solid angle
and trails behind the supernova debris.
Because the supernova flow is supersonic, the hole in the debris
will not close with time. However,
the orbital motion of the main sequence and subgiant secondaries will add
a small component of velocity ($100-200$ \kmsecc)
tangential to the axis of the
original binary system that will slowly distort the shape
of the slow-moving contaminated region behind most of the
supernova debris. The evolution and mixing of the hydrogen-filled region 
become a 3-D problem.

We use the angular size of the hole
as an estimate of the region of asymmetry in the supernova ejecta
caused by the impact on the companion star.
The angular size of the gap in the supernova debris depends
on the binary scenario. A typical angular size for the gap in the 
high-velocity material is $31^{\circ}- 34^{\circ}$, which corresponds to
only $7-8\%$ of the ejecta's surface. The main sequence model
has the smallest gap, the subgiant has a slightly larger gap, 
and the red giant has the largest gap. The typical angular size
for the gap in the low-velocity ejecta is $\sim 40^{\circ}$, 
independent of the model. This corresponds to $12\%$ of the 
surface area of the ejecta. Because the binary scenarios we
explore are all close enough, or almost close enough, to
be in Roche lobe overflow, the degree of asymmetry is
similar for all the models. 
Asymmetry in the supernova ejecta will indicate the presence
of a companion star, but it will be difficult to use the degree of
asymmetry alone to discriminate between main sequence, subgiant, and
red giant companions.

The asymmetry in the supernova debris could have observational
consequences beyond the change in morphology of the supernova 
remnant. In the early-time spectra, subtle distortions may appear 
in the P Cygni lines that are dependent on the orientation of
the hole in the supernova debris relative to the observer.
For example, if the gap lies towards the observer, the P Cygni 
lines may have a distorted blue wing. If the gap lies at some
angle away from the observer, the P Cygni lines may have a
distorted red wing. However, because the hole in the supernova debris 
subtends such a small solid angle, its effect on the spectra will be equally small. 
The asymmetry may also affect the polarization of the supernova debris 
to some degree.

The impact of the supernova shell will also have consequences for the
future evolution of secondary star. After the impact, the main sequence star
is puffed up, much like a pre-main sequence star. 
Although the energy generation rate during the impact shoots up by a factor of 
$\sim1000$, it immediately plummets as the star expands and cools. 
Hydrogen burning in the core will be nearly extinguished, but
the luminosity will rise dramatically, to as high as $\sim5000$ \Lsunn,
as the extended envelope relaxes back into thermal equilibrium 
and then follows a \kh track on a timescale of $\sim10^{4}$ yrs. 
The supernova is initially much more
luminous with L$\sim 10^{9}$ \Lsunn. However, after $2-3$ years,
the supernova luminosity will have decayed enough that a very bright
secondary might be visible.
After thermal equilibrium is reestablished and the residue has
settled back onto the main sequence, the remnant will
continue its evolution at a rate prescribed by its new mass.
The $1.0$ \Msun main sequence secondary will eventually return to the main 
sequence with a mass of $0.85$ \Msunn, having lost $0.15$ \Msun due to the impact, 
with a slightly smaller radius of $0.9$ \Rsunn,
and a luminosity of $\sim0.4$ \Lsunn.
The subgiant puffs up after the impact and must contract to 
reignite hydrogen shell-burning, which has been extinguished
during the subgiant's expansion. Like the main sequence case, the luminosity will be 
high during this phase.

A He pre-white dwarf will be left after a supernova explosion
with a red giant companion. Almost all the envelope will
be ejected by the impact, but a residual amount of material
will form an extended, hydrogen-rich envelope around the degenerate core.
This extended envelope of $\sim 0.02$ \Msun of hydrogen-rich material
will settle down onto the star to feed the hydrogen burning shell.
The star will evolve away from the red giant branch at constant
luminosity and increasing effective temperature on a timescale of
$10^{5}-10^{6}$ years. Before the luminosity drops,
it may appear as an underluminous O or B star. 
When the star begins to dim and cool, it may pass
through an sdO or sdB phase before retiring
along a standard He white dwarf cooling track.
This binary scenario could be a possible pathway for the formation
of a subset of single, low-mass He white dwarfs.

The metallicity of the secondary may increase if it accretes any
supernova material as a result of the explosion.
We do not expect that the secondary star will accrete a significant amount of 
supernova material by the direct impact of the supernova shell on the surface 
of the star because, as the shock created by the impact
propagates through the stellar envelope, the shock-heated material
escapes the secondary, carrying away any impinging supernova material.
However, this does not necessarily mean that the secondary will not
accrete any low-velocity iron-group elements (or oxygen and silicon if the
ejecta is radially mixed) long after the initial impact.

High resolution late-time observations focused on 
detecting stripped hydrogen and helium from the companion
star are needed to
 verify the single-degenerate
scenario. With non-LTE synthetic spectra, such observations
may be used to discriminate between the main sequence,
the subgiant, and the red giant companions. 
Asymmetry in the supernova ejecta, distorted P Cygni lines or line ratios, 
and slight polarization may also indicate an unseen companion star.
Our theoretical study was performed in support of the quest
to unravel the persistent mystery of the site of Type Ia supernova
explosions. Given the emerging cosmological importance of such expolosions
as standardizable candles and the resulting renaissance in their study, the time
is ripe for a renewed investigation into the consequences of
the variety of binary scenarios for thermonuclear supernovae.

\acknowledgements{ 
We thank Brian Chaboyer for the stellar models used in the simulations
and Matthias Steinmetz for the Poisson solver.
Conversations with Phil Pinto, Jim Liebert, Peter Hoeflich, and Betsy Green are gratefully
acknowledged.
This work was supported by NSF grant AST96-17494,
NASA GSRP Fellowship NGT-51305,
and by the ASCI Center for Thermonuclear Flashes
at the University of Chicago, under DOE contract B341495.
The calculations were performed on the Cray C90 at Pittsburgh
Supercomputer Center and on the Cray T90 at the San Diego
Supercomputer Center.
Postscript images, MPEG movies, and a selection of figures
presented in this paper, are posted at http://www.astrophysics.arizona.edu and are
available via FTP at www.astrophysics.arizona.edu, in directory pub/marietta.
This research has made use of NASA's Astrophysics Data System Abstract Service.

{}

\newpage


\figcaption{ Mass vs.~radius for the
main sequence, subgiant, and red giant secondaries.
 \label{m} }

\figcaption{ Absolute value of the exterior binding 
energy of the main sequence, subgiant, and red giant secondaries. 
The binding energy of each secondary is provided next to
its legend. The value listed for the red giant secondaries
is the binding energy of the envelope only. The dashed line indicates
the binding energy profile of the red giant core.
\label{be} }

\figcaption{ Frames (a)-(d) illustrate
the impact of supernova Type Ia W7 on a $1.0$ \Msun main sequence
companion at $29$, $104$, $179$ and $329$ seconds after the explosion.
The color scale indicates Log$_{10} \rho$:
blue $\sim 160.0$ gm cm$^{-3}$ $\rightarrow$
red $\sim 10^{-13}$ gm cm$^{-3}$.
The explosion occurs off the grid at a distance consistent
with the companion losing mass to the white dwarf primary
by Roche lobe overflow. 
Frame (a) shows the companion star
at the origin of the grid at $29$ seconds, just prior to the 
impact of the debris.
In Frame (b), at $104$ seconds,  the leading edge of the 
supernova shell has just collided with the surface of the companion.
By $179$ seconds (Frame (c)), the leading edge of the
supernova shell converges along the downstream axis.
The companion is now enveloped in the supernova ejecta.
Frame (d), at $329$ seconds, shows the shock wave that
was driven into the companion by the impact and the
bow shock just upstream of the companion.
MPEG movies and  2-D images are
posted at http://www.astrophysics.arizona.edu and are available via FTP
from www.astrophysics.arizona.edu, in directory pub/marietta.
\label{ms1} }

\figcaption{ Frames (a)-(d) illustrate the impact of supernova 
Type Ia W7 on a $1.0$ \Msun main sequence companion
at $529$, $755$, $2029$ and $2033$ seconds after the explosion.
The color scale is the same as in Figure \ref{ms1}.
In Frame (a) the shock has just passed through the stellar core.
In Frame (b) the shock has converged within the star 
along the downstream axis.
\kh and \rt instabilities are visible in both frames.
Frame (c) illustrates the end of the mass stripping phase.
By Frame (d), several dynamical times have passed, and the
remnant is beginning to recover hydrostatic equilibrium.
\label{ms2} }

\figcaption{ Frames (a)-(f) illustrate the 
impact of supernova Type Ia W7 on a $1.0$ \Msun 
main sequence companion in a series of cartoon images, 
each color indicating the dominant element in that region.
For clarity each region is also numbered: (1) circumstellar
medium, (2) stellar hydrogen, (3) supernova oxygen-group
elements, (4) supernova silicon, and (5) supernova
iron-group elements. 
The frames shown are 
at $104$, $179$, $329$, $529$, $755$, and $2029$ seconds 
after the explosion.
Frames (a) and (b) show the initial impact. 
The oxygen, silicon, and iron layers of the supernova ejecta
are clearly visible. 
Frames (c) and (d) illustrate the beginning of
the mass stripping phase when the outermost layer of the
companion's envelope is stripped and caught in the
oxygen and silicon layers of the supernova ejecta.
By Frame (e), any mass lost from the companion's 
envelope is embedded within the iron region of the supernova ejecta.
Frame (f) shows the end of the mass stripping phase.
The envelope of the companion is now extended; about
$15\%$ of the mass will escape the companion and
follow the flow of the supernova.
\label{ms3} }

\figcaption{
Momentum and density profile of SNIa W7 and SNIa Hedt.
SNIa W7 is shown at $24$ hrs while SNIa Hedt is shown
at $27.87$ hrs.
\label{SN} }

\figcaption{Pressure profile of the initial impact
on the main sequence secondary (HCV) along the z axis,
illustrating the propagation of the shock through the stellar
core and the development of the bow shock. 
The profiles shown are at 
$29$, $104$, $179$, $257$, $332$, $451$,
$601$, $676$, and $2030$ seconds after the explosion and
are labeled 1-9, respectively.
The first and last timesteps are marked by
thicker lines. They represent the 
undisturbed companion and the companion after the
mass stripping is complete and are included for reference.
\label{press1_ms} }

\figcaption{Pressure profile
of the main sequence secondary (HCV) along the z axis,
illustrating the abrupt pressure rise in the back of the
star as the stellar shock converges along the axis.
As the shock-heated material is ejected from the
star, the back pressure drops. 
The profiles shown are at
$29$, $751$, $827$, $902$, $1026$, $1178$, $1528$,
and $2030$ seconds after the explosion and are labeled
1-8, respectively.
The first and last timestep are marked by
thicker lines. They represent the 
undisturbed companion and the companion after the
mass stripping is complete.
\label{press2_ms} }

\figcaption{Pressure profile of the
main sequence secondary (HCV) along the z axis
as the star recovers hydrostatic equilibrium.
The remnant is now moving down the axis 
with the kick received by the impact.
The profiles shown are at 
$29$, $2030$, $5035$, $10053$, $15056$, and $20032$
seconds after the explosion and are labeled
1-6, respectively.
The first and last timesteps are marked by
thicker lines; they represent the
undisturbed companion and the post-impact 
companion after several pulsations.
\label{press3_ms} }

\figcaption{Central density and temperature
as a function of time for the HCV scenario 
($1.0$ \Msun secondary)
\label{DTts98}}

\figcaption{Pre- and Post-impact density 
profile of the the main sequence secondary (HCV)
and the subgiant secondary (HCVL).
\label{Dens1} }

\figcaption{Pre- and Post-impact entropy
profile (per Boltzmann's constant per baryon) 
of the main sequence secondary (HCV) 
and the subgiant secondary (HCVL). 
\label{S1} }

\figcaption{Estimate of nuclear energy generation 
from the pp chains and the CNO cycle for the HCV scenario
($1.0$ \Msun secondary) using a 1-D section with
spherical symmetry. The compression and expansion of
the stellar core releases $\sim 10^{38}$ ergs in 
the first $1000$ seconds. 
\label{NucTs98} }

\figcaption{ 
A. Velocity distribution $ dM/ dV $ of 
the stripped stellar hydrogen and helium for 
the HCV binary scenario
at the end of the simulation, along with the
velocity distribution of SNIa W7.
The velocity at the half-mass point of the HCV distribution, 
as indicated by the vertical line and the arrow,
is $823$ \kmsecc. In contrast, the
velocity at the half-mass point of the SN W7 distribution, 
also indicated in the plot, is
$7836$ \kmsecc.
B. Velocity distribution $dM/dV$ of 
the stripped stellar hydrogen and helium for 
the HCVL binary scenario
at the end of the simulation, along with the
velocity distribution of SNIa W7.
The velocity at the half-mass point of the HCVL distribution, 
as indicated by the vertical line and the arrow,
is $887$ \kmsecc. 
\label{dmdv_bw2} }

\figcaption{ 
The solid angle distribution of the stripped envelope
of the main sequence and subgiant simulations HCV and HCVL.
The half-mass angle of each distribution is listed.
The dashed lines indicate the distribution of supernova
material for both simulations.
The hole in the supernova debris created by the
impact is visible between $0^{\circ}-40^{\circ}$.
\label{dmds_ts85_ts88} }

\figcaption{
The magnitude of the angular distribution (dM/d$\alpha$)
is plotted as the radial coordinate for the HCV and HCVL
binary scenarios. The angular coordinate is the angle
from the downstream axis. The geometrical shadow is
shown for both simulations. 
\label{polar_ts85_ts88}}

\figcaption{ 
Hydrogen mass fraction of the
contaminated supernova ejecta for the HCV scenario 
($1.0$ \Msun main sequence secondary) 
vs.~the mass of the original
W7 supernova model. 
The solid line
indicates the hydrogen mass fraction assuming that the
stripped material is uniformly mixed in angle with 
Type Ia model W7.
The mass fractions of the O, Si, and Fe-group
elements are shown for comparison.
Because the stripped material in this simulation
contaminates only the supernova ejecta within $65.7^{\circ}$ 
of the downstream axis, the dashed line indicates
the hydrogen mass fraction assuming mixing only in
this region.
\label{Mts85} }

\figcaption{ 
Hydrogen mass fraction of the
contaminated supernova ejecta for the HCV scenario 
($1.0$ \Msun main sequence secondary) 
vs.~velocity. See the caption for Figure \ref{Mts85}.
\label{Vts85} }

\figcaption{ Frames (a)-(d) illustrate the 
impact of supernova Type Ia W7 on a $1.0$ \Msun 
main sequence companion for different binary
separations: simulations HCV, HCVb, HCVc, and
HCVd. 
The color scale indicates Log$_{10} \rho$:
blue $\sim 160.0$ gm cm$^{-3}$ $\rightarrow$
red $\sim 10^{-13}$ gm cm$^{-3}$.
This color scale is the same as in Figure \ref{ms1}.
Frame (a) shows the impact for
a ratio $a/R = 3.0$ (binary separation over stellar radius)
at $1031$ seconds after the explosion.
This is our HCV simulation.
Frame (b) shows the impact with $a/R = 4.0$ 
at $1336$ seconds after explosion, the
HCVb test case.
Frame (c) shows the impact with $a/R = 6.0$
at $1197$ seconds, and Frame (d) with
$a/R = 12.0$ at $1308$ seconds.
The timesteps are increasing primarily because
the length of time needed for the supernova
ejecta to reach the secondary is increasing.
\label{ms1_em5} }

\figcaption{ 
The magnitude of the angular distribution  (dM/d$\alpha$)
is plotted as the radial coordinate for the 
main sequence simulation HCV and 
test cases HCVa, HCVb, and HCVc,
which differ only in their distance from the supernova explosion.
The half-mass angles of the distributions are 
$45.1^{\circ}$, $42.7^{\circ}$, $35.8^{\circ}$, $24.7^{\circ}$,
and $9.9^{\circ}$ for the HCVa, HCV, HCVb, HCVc, and HCVd 
simulations, respectively.
\label{polar_em5series} }

\figcaption{ Frames (a)-(d) illustrate the 
impact of supernova Type Ia W7 on a $1.13$ \Msun 
subgiant companion 
at $343$, $945$, $2044$ and $20044$ seconds after the explosion.
The color scale indicates Log$_{10} \rho$:
blue $\sim 2.7\times10^{3}$ gm cm$^{-3}$ $\rightarrow$
red $\sim 10^{-13}$ gm cm$^{-3}$.
The explosion occurs off the grid at a distance consistent
with the companion losing mass to the white dwarf primary
by Roche lobe overflow. 
Frame (a) shows the subgiant and the incoming supernova
shell $343$ seconds after the explosion.
Only $0.06$ \Msun of the companion's envelope has
been stripped by this time.
In Frame (b) at $945$ seconds, the shock has just propagated
through the subgiant's core and is about to converge in the
back of the subgiant.
Frame (c) shows the subgiant in the last stages of mass loss.
The subgiant has lost $0.15$ \Msun of its envelope which can
be seen streaming away from the subgiant.
By Frame (d), $20044$ seconds after the explosion,
the remnant of the subgiant is recovering hydrostatic equilibrium.
\label{sg1} }

\figcaption{ Impact of supernova Type Ia W7 on a $1.13$ \Msun 
subgiant companion at $1958$ seconds near the end of the mass stripping
phase. The \kh instabilities, which help the hot material to
escape the surface, are easily seen in this image.
The color scale is the same as in Figure \ref{sg1}.
\label{sg2} }

\figcaption{
Hydrogen mass fraction of the
contaminated supernova ejecta for the HCVL scenario 
($1.1$ \Msun subgiant secondary) 
vs.~the mass of the original
W7 supernova model. 
The solid line
indicates the hydrogen mass fraction assuming that the
stripped material is uniformly mixed in angle with W7.
The mass fractions of the O, Si, and Fe-group
elements are shown for comparison.
Because the stripped material in this simulation
contaminates the supernova ejecta within only $72.5^{\circ}$ 
of the downstream axis, the dashed line indicates
the hydrogen mass fraction assuming mixing only in
this region.
\label{Mts88} }

\figcaption{
Hydrogen mass fraction of the
contaminated supernova ejecta for the HCVL scenario 
($1.1$ \Msun subgiant secondary) 
vs.~velocity. 
See the caption for Figure \ref{Mts88}.
\label{Vts88} }

\figcaption{ Frames (a)-(d) illustrate the 
impact of supernova Type Ia Hedt on a $0.98$ \Msun 
red giant companion at $8.7$ hrs, $14.3$ hrs, 
$2.1$ days and $5.9$ days after the explosion. 
Only the region around the red giant is shown
to emphasize the interaction. The full images
are shown in Figures \ref{rg2}.
\label{rg1} }

\figcaption{ Frames (a)-(d) illustrate the 
impact of supernova Type Ia Hedt on a $0.98$ \Msun 
red giant companion at $8.7$ hrs, $14.3$ hrs, 
$2.1$ days and $5.9$ days after the explosion.
In these images the supernova can been seen exploding on the grid.
These are the full images corresponding to
the images in Figures \ref{rg1}. 
\label{rg2} }

\figcaption{ 
A. Velocity distribution $dM/dV$ of 
the stripped stellar hydrogen and helium for 
the HALGOLa binary scenario
at the end of the simulation, along with the
velocity distribution of SNIa W7.
The velocity at the half-mass point of the HALGOLa distribution, 
as indicated by the vertical line and the arrow,
is $593$ \kmsecc. In contrast, the
velocity at the half-mass point of the SN W7 distribution, 
also indicated in the plot, is
$7836$ \kmsecc.
B. Velocity distribution $dM/dV$ of 
the stripped stellar hydrogen and helium for 
the SYMB binary scenario.
The velocity at the half-mass point of the SYMB distribution, 
as indicated by the vertical line and the arrow,
is $421$ \kmsecc. The velocity at the half-mass point of 
the SN Hedt distribution, also indicated in the plot, is
$10000$ \kmsecc.
\label{dmdv_bw1} }

\figcaption{ 
Solid angle distribution of the stripped envelope
of the red giant simulations HALGOLa and SYMB.
The dashed lines indicate the distribution of supernova material.
The gap between $0^{\circ}-40^{\circ}$ illustrates the
hole in the supernova ejecta created by the impact.
\label{dmds_t117_t120_sn} }

\figcaption{ 
The magnitude of the angular distribution (dM/d$\alpha$)
is plotted as the radial coordinate for the red giant
simulations (HALGOLa and SYMB).
\label{polar_t117_t120} }

\figcaption{
Hydrogen mass fraction of the
contaminated supernova ejecta for the HALGOLa scenario 
($0.98$ \Msun red giant secondary) 
vs.~the mass of the original
W7 supernova model. 
The solid line
indicates the hydrogen mass fraction assuming that the
stripped material is uniformly mixed in angle with Type Ia model W7.
The mass fractions of the O, Si, and Fe-group
elements are shown for comparison.
Because the stripped material in this simulation
contaminates the supernova ejecta within only $115.4^{\circ}$ 
of the downstream axis, the dashed line indicates
the hydrogen mass fraction assuming mixing only in
this region.
\label{Mt117} }

\figcaption{
Hydrogen mass fraction of the
contaminated supernova ejecta for the HALGOLa scenario 
($0.98$ \Msun subgiant secondary) 
vs.~velocity. 
See the caption for Figure \ref{Mt117}.
\label{Vt117} }

\figcaption{ Mass of stripped hydrogen
with velocity $\ge$ V for the HCV ($1.0$ main sequence),
HCVL ($1.13$ \Msun subgiant), SYMB and HALGOLa 
($1.0$ \Msun red giant) simulations.  
To compare the mass profiles to Type Ia supernova
observations near maximum light,
we indicate with a horizontal line the Della Valle \etal (1996)
upper limit of $3\times10^{-4}$ \Msun of hydrogen
in the spectrum of SN 1990M.
This assumes perfect radial mixing of the silicon and hydrogen.
Because Della Valle \etal (1996) estimated
the error in the upper limit to be a factor of $2-3$,
we also include horizontal lines to indicate this range.
Della Valle \etal (1996) estimated that the size of the photosphere
was $0.04$ \Msun at the time of the observation. The velocity
corresponding to this mass is estimated from SNIa W7 and shown
in the figure as a vertical line.
\label{highv_bi}}

\begin{center}

 \begin{deluxetable}{l l l}
 \tablewidth{6in}
 \tablecaption{Single Degenerate Type Ia Models \label{singlemodels} }
 \tablehead{
 \colhead{Single Degenerate SNIa} & \colhead{Secondary}&
 \colhead{Mass Transfer} }
 \startdata
Hydrogen Cataclysmic Variables     & Main Sequence/Subgiant &
     RLOF, H \nl
Hydrogen Cataclysmic-Like Variables & Main Sequence/Subgiant &
     RLOF, H \nl
                                &                        &  \nl

Symbiotic Stars                 & Red Giant              &
     Wind, H \nl
Hydrogen Algols                 & Red Giant              &
     RLOF, H \nl
                                &                        &  \nl

Helium Cataclysmic Variables        & He Star                &
     RLOF, He   \nl

Helium Algols                       & He Giant              & 
    RLOF, He    \nl

 \enddata
 \end{deluxetable}

 \begin{deluxetable}{ l l l l l }
 \tablewidth{6in}
 \tablecaption{Secondary Models \label{secondarymodels} }
 \tablehead{
	 \colhead{Evolutionary Stage}         & 
	 \colhead{M (\Msunn)}                 & 
	 \colhead{R (km)}                     & 
	 \colhead{$t_{dyn}$   (s) \tablenotemark{a} } & 
         \colhead{$t_{cross}$ (s) \tablenotemark{b} } 
 }
 \startdata
           Main Sequence           & 
           1.017                   &
           $6.80\times10^{5}$      &
           $1.5\times10^{3}$       &
           $2.6\times10^{3}$       \nl

 	   Subgiant                 &
            2.118                   &
            $1.62\times10^{6}$      &
            $4.0\times10^{3}$       &
            $5.8\times10^{3}$      \nl

	   Subgiant, mass loss     &
           1.132                   &
           $1.22\times10^{6}$      &
           $3.5\times10^{3}$       &
           $5.5\times10^{3}$       \nl

           Red Giant \tablenotemark{c}  &
           0.977                   &
           $1.19\times10^{8}$      &
           $6.4\times10^{6}$       &
           $5.5\times10^{6}$       \nl

 \enddata
\footnotesize
\tablenotetext{a}{Dynamical time, 
                  $t_{dyn} = \frac{1}{2} \left(  G \langle \rho \rangle \right)^{-1/2}$ }
\tablenotetext{b}{Sound crossing time, $t_{cross} = R / \langle c_{s} \rangle $, integrated over star}
\tablenotetext{c}{Softened $\phi(r) = -G M_c / ( r + r_{c} )$, 
                  $M_{c} = 0.424$ \Msunn, $R_{c} = 5 \times 10^{5}$ km }
\end{deluxetable}

 \begin{deluxetable}{ l l l l l l }
 \tablewidth{6in}
 \tablecaption{Supernova Ejecta Models \label{ejectamodels} }
 \tablehead{
         \colhead{Model}                                      & 
	 \colhead{M (\Msunn)}                                & 
	 \colhead{E ($10^{51}$ ergs)}                         & 
	 \colhead{P (\Msun $V_{c}$)   \tablenotemark{a} }     & 
          \colhead{$ \langle V \rangle$ \tablenotemark{b} }   &
          \colhead{V$_{{1 \over 2} M}$ \tablenotemark{c} }
}
 \startdata
	   \SNIa W7 \tablenotemark{d} &
           1.378                   & 
           1.233                   &
           1.175                   &
          $8.527 \times 10^{3}$    &
          $7.836 \times 10^{3}$    \nl

	   \SNIa Hedtb11  \tablenotemark{e} &
           0.899                   & 
           1.079                   &   
           0.912                   &
          $1.014 \times 10^{4}$    &
          $1.006 \times 10^{4}$   \nl

 \enddata
\footnotesize
\tablenotetext{a}{Momentum of supernova ejecta, specified in units of
                  \Msun and divided by V$_{c} = 10^9$ cm s$^{-1}$ }
\tablenotetext{b}{Average speed of the supernova ejecta, km s$^{-1}$}
\tablenotetext{c}{Speed of the supernova ejecta at the half-mass point, km s$^{-1}$}
\tablenotetext{d}{\cite{nomoto84}}
\tablenotetext{e}{\cite{woosley94}}
\end{deluxetable}
\clearpage

 \begin{deluxetable}{l l l l r l }
 \tablewidth{6in}
 \tablecaption{Simulations \label{sims} }
 \tablehead{
	 \colhead{Simulations}     & 
	 \colhead{   }             & 
	 \colhead{M (\Msun)}       & 
         \colhead{a (km)}          & 
         \colhead{a/R}             & 
         \colhead{SN Model} 
 }
 \startdata
HCV                                &
                                   &
           1.017 (MS)              &
           $2.04\times10^{6}$      &
           3.00                    &
           W7                      \nl

& \multicolumn{5}{l}{Additional Separations} \nl

                                   &
HCVa                               &
           1.017 (MS)              &
           $1.75\times10^{6}$      &
           2.57                    &
           W7                      \nl

                                   &
HCVb                               & 
           1.017 (MS)              &
           $2.72\times10^{6}$      &
           4.00                    &
           W7                      \nl

                                   &
HCVc                               & 
           1.017 (MS)              &
           $4.08\times10^{6}$      &
           6.00                    &
           W7                      \nl

                                   &
HCVd                               &
           1.017 (MS)              &
           $8.16\times10^{6}$      &
           12.00                   &
           W7                      \nl

HCVL                               &
                                   &
           1.132  (SG)             &
           $3.39\times10^{6}$      &
           2.78                    &
           W7                      \nl

& \multicolumn{5}{l}{No Mass Loss} \nl
     
                                   &
HCVLa                              &
           2.118  (SG)             &
           $4.50\times10^{6}$      &
           2.78                    &
           W7                      \nl

HALGOLa                            &
                                   &
           0.977 (Red Giant)       &
           $3.00\times10^{8}$      &
           2.52                    &
           W7                      \nl

SYMB                               &
                                   &
           0.977 (Red Giant)       &
           $3.76\times10^{8}$      &
           3.16                    &
           Hedt                    \nl

 \enddata
\end{deluxetable}

 \begin{deluxetable}{l l l l}
 \tablewidth{6in}
 \tablecaption{Pre- and Post-Impact Central Values for HCV Companion \label{postimpactchart}}
 \tablehead{
	 \colhead{   }     & 
         \colhead{Initial} &
         \colhead{Final}   &
          \colhead{Fractional Change}
 }
 \startdata
          Pressure (dynes cm$^{-2}$)  &
          $2.34\times10^{17}$    &
          $9.41\times10^{16}$    &
          $0.60$                \nl

          Density (gm cm$^{-3}$) &
          $1.54\times10^{2}$     &
          $7.95\times10^{1}$     &
          $0.48$                \nl

          Temperature (K)        &
          $1.56\times10^{7}$     &
          $9.68\times10^{6}$     &
          $0.38$                \nl

          Entropy (k$_{b}^{-1}$ baryon$^{-1}$) &
          $1.10\times10^{1}$     &
          $1.35\times10^{1}$     &
          $0.23$                \nl

\footnotesize
 \enddata
 \end{deluxetable}

 \begin{deluxetable}{l r  l l }
 \tablewidth{6in}
 \tablecaption{ Stripped Mass\label{stripchart} }
 \tablehead{
	 \colhead{Scenario}                              & 
         \colhead{Time (s)}                              & 
         \colhead{$\Delta M$ (\Msunn)\tablenotemark{a}}  &
         \colhead{$\Delta M$ (\Msunn)\tablenotemark{b}}            
 }
 \startdata

HCV                              &
            $2.0\times10^{4}$     & 
            0.15                  &
            0.16                  \nl

HCVa	                          &
            $2.0\times10^{4}$     & 
            0.23                  &
            0.20                 \nl
HCVb	                          &
            $2.5\times10^{4}$     & 
            0.074                 &
            0.11                 \nl
HCVc	                          &
            $3.0\times10^{4}$     & 
	    0.022                 &
            0.063                 \nl
HCVd	                          &
            $3.0\times10^{4}$     &
            0.0018                &
            0.021                \nl

HCVL	                          &
            $3.0\times10^{4}$     & 
            0.17                  &
            0.14                 \nl
HCVLa	                          &
            $3.0\times10^{4}$     & 
            0.25                  &
            0.21                 \nl

HALGOLa	                  &
             $6.0\times10^{6}$     & 
             0.54 (98\%)           &
             0.55 (100\%)          \nl
SYMB	                          &
             $8.0\times10^{6}$     & 
             0.53 (96\%)           &
             0.55 (100\%)          \nl

 \enddata
\footnotesize
\tablenotetext{a}{Numerical Calculation of Stripped Mass}
\tablenotetext{b}{Analytic Estimate of Stripped Mass, method of
\cite{wheeler75}}
 \end{deluxetable}

 \begin{deluxetable}{l r r r r r}
 \tablewidth{6in}
 \tablecaption{Kicks \label{kickchart} }
 \tablehead{
	 \colhead{Scenario}                                    & 
         \colhead{M$_{rem}$ (\Msunn) \tablenotemark{a}}        &
          \colhead{V$_{rem}$ \tablenotemark{b}}                &
          \colhead{V \tablenotemark{c}}                        &
          \colhead{V$_{orb}$ \tablenotemark{d}}                &
         \colhead{ P$_{rem}$/P$_{in}$ \tablenotemark{e}}
 }
 \startdata

HCV                              &
            0.867                 &
             85.7                 &
            109.8                 &
            227.1                 &
            0.228                 \nl

HCVa	                          &
           0.787                  &
             99.2                 &
            137.1                 &
            245.2                 &
            0.176                 \nl

HCVb	                          &
           0.943                  &
            61.5                  &
            73.2                  &
           196.7                  &  
           0.316                 \nl

HCVc	                          &
           0.995                  &
            35.2                  &
            41.7                  &
           160.6                  &
           0.429                 \nl

HCVd	                          &
           1.015                  &
            10.1                  &
            14.8                  &
           113.5                  &
           0.503                 \nl

HCVL	                          &
               0.962              &
              49.4                &
              70.4                &
             172.1                &
             0.125                \nl

HCVLa	                          &
            1.868                 &
            54.4                  &
            65.4                  &
           126.5                  &
           0.267                 \nl
 \enddata
\footnotesize
\tablenotetext{a}{Numerical Calculation of the Mass of the Stellar Remnant}
\tablenotetext{b}{Numerical Calculation of the Kick of the Stellar Remnant, \kmsecc}
\tablenotetext{c}{Ballpark Estimate of Kick from Inelastic Collision, \kmsecc}
\tablenotetext{d}{Orbital Velocity, \kmsecc}
\tablenotetext{e}{Ratio of Stellar Remnant Momentum (P$_{rem}$) to Incident Supernova 
                  Momentum (P$_{in}$)}
 \end{deluxetable}

 \begin{deluxetable}{l l l l l }
 \tablewidth{6in}
 \tablecaption{Mass of Stripped Hydrogen with Velocity $\ge$ V at
  four chosen points: 
    V$_{A} = 3.0\times10^{3}$ km s$^{-1}$,
    V$_{B} = 1.0\times10^{4}$ km s$^{-1}$,
    V$_{C} = 1.2\times10^{4}$ km s$^{-1}$, and
    V$_{D} = 1.5\times10^{4}$ km s$^{-1}$.
 \label{highvchart} }
 \tablehead{
	 \colhead{Scenario}   & 
         \colhead{ A }  & 
         \colhead{ B }  & 
         \colhead{ C }  & 
         \colhead{ D }
 }
 \startdata

HCV                                   &
             $9.7\times10^{-3}$ \Msun &
             $1.3\times10^{-3}$ \Msun &
             $6.9\times10^{-4}$ \Msun &
             $1.3\times10^{-4}$ \Msun \nl

HCVL	                              &
             $1.4\times10^{-2}$ \Msun &
             $1.4\times10^{-3}$ \Msun &
             $7.8\times10^{-4}$ \Msun &
             $1.7\times10^{-4}$ \Msun \nl

HALGOLa	                              &
             $1.2\times10^{-2}$ \Msun &
             $1.4\times10^{-3}$ \Msun &
             $8.9\times10^{-4}$ \Msun &
             $3.3\times10^{-4}$ \Msun \nl

SYMB	                              &
             $4.3\times10^{-3}$ \Msun &
             $6.2\times10^{-4}$ \Msun &
             $4.8\times10^{-4}$ \Msun &
             $3.0\times10^{-4}$ \Msun \nl
 \enddata
 \end{deluxetable}

\end{center}

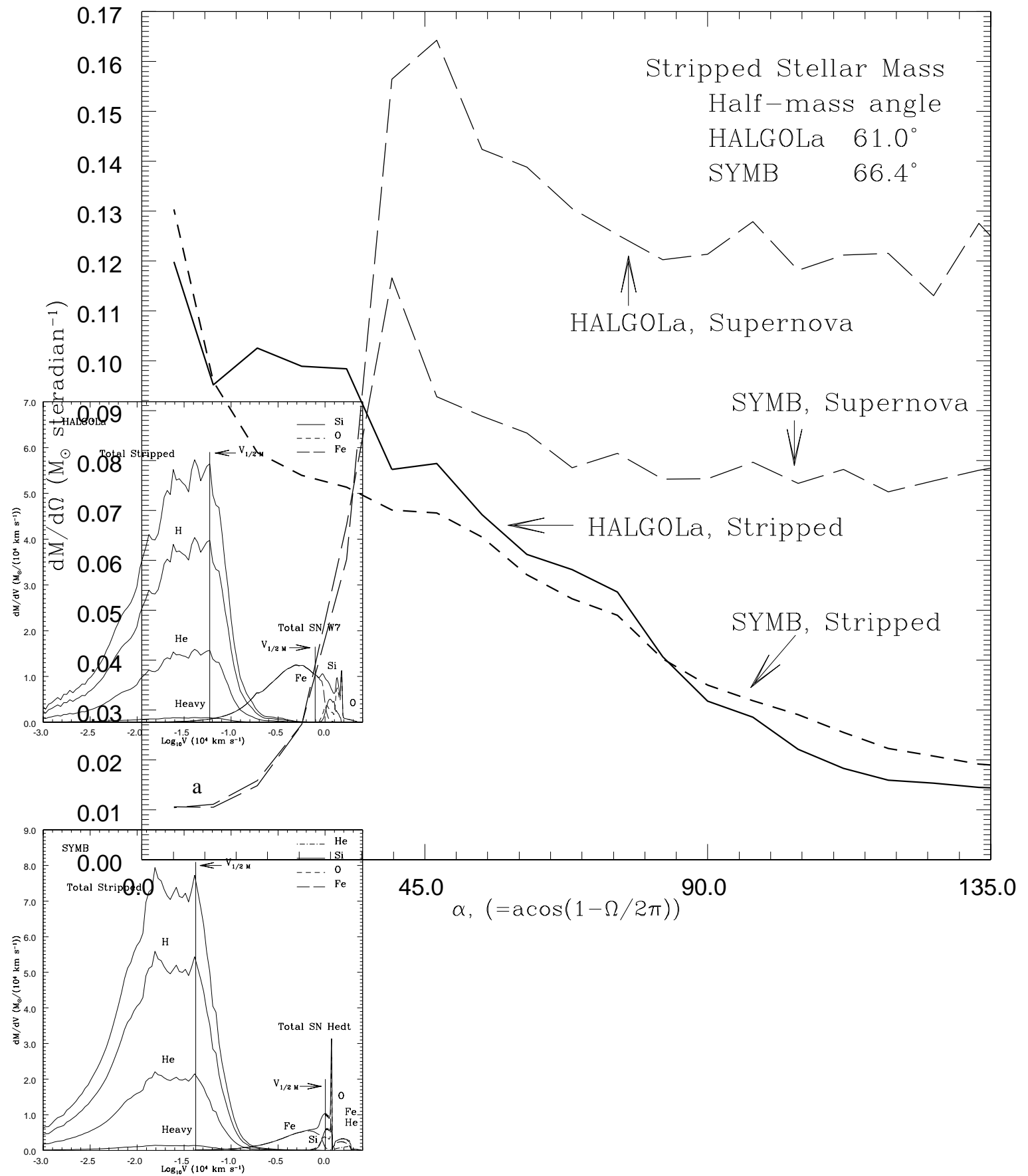
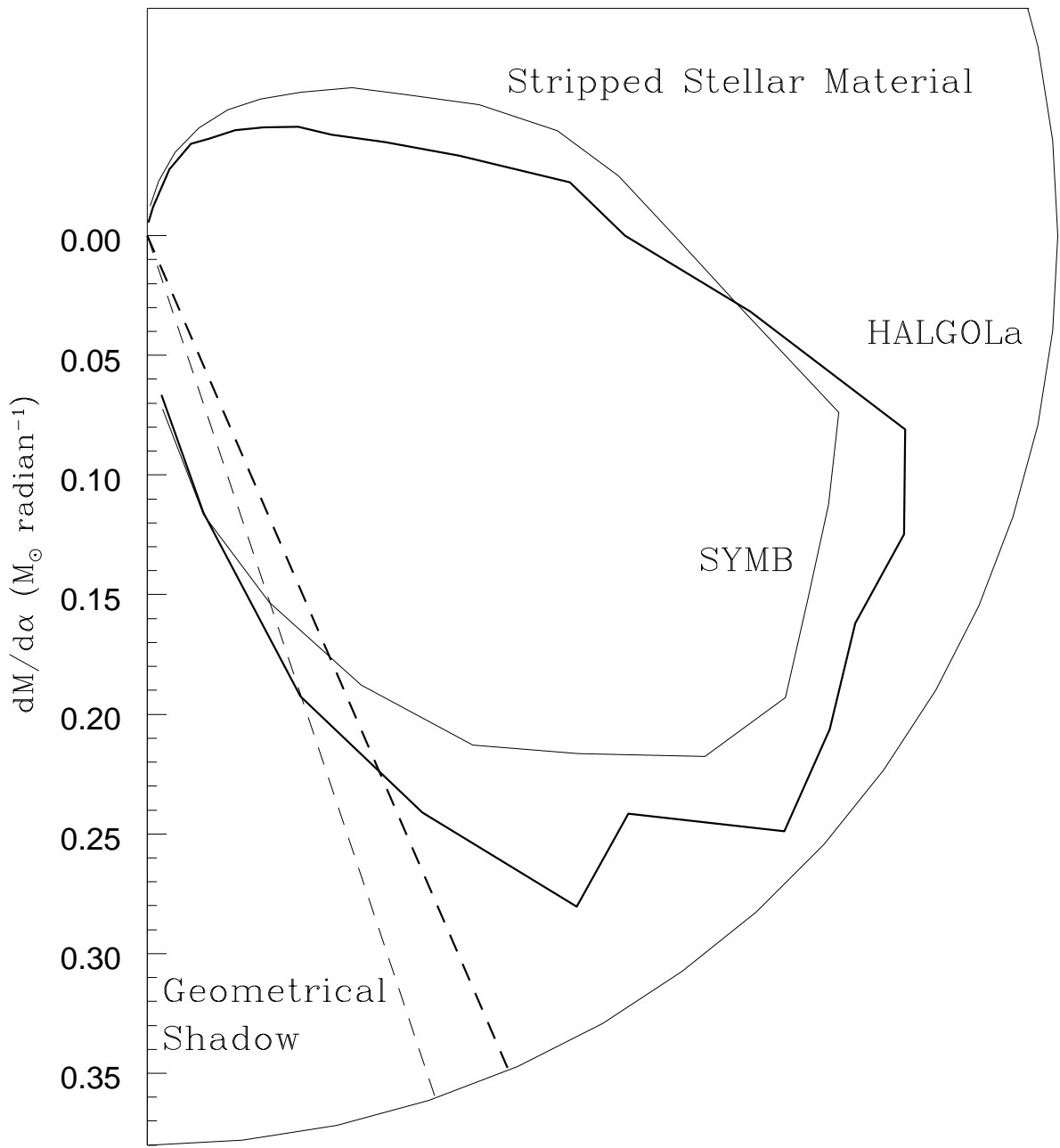
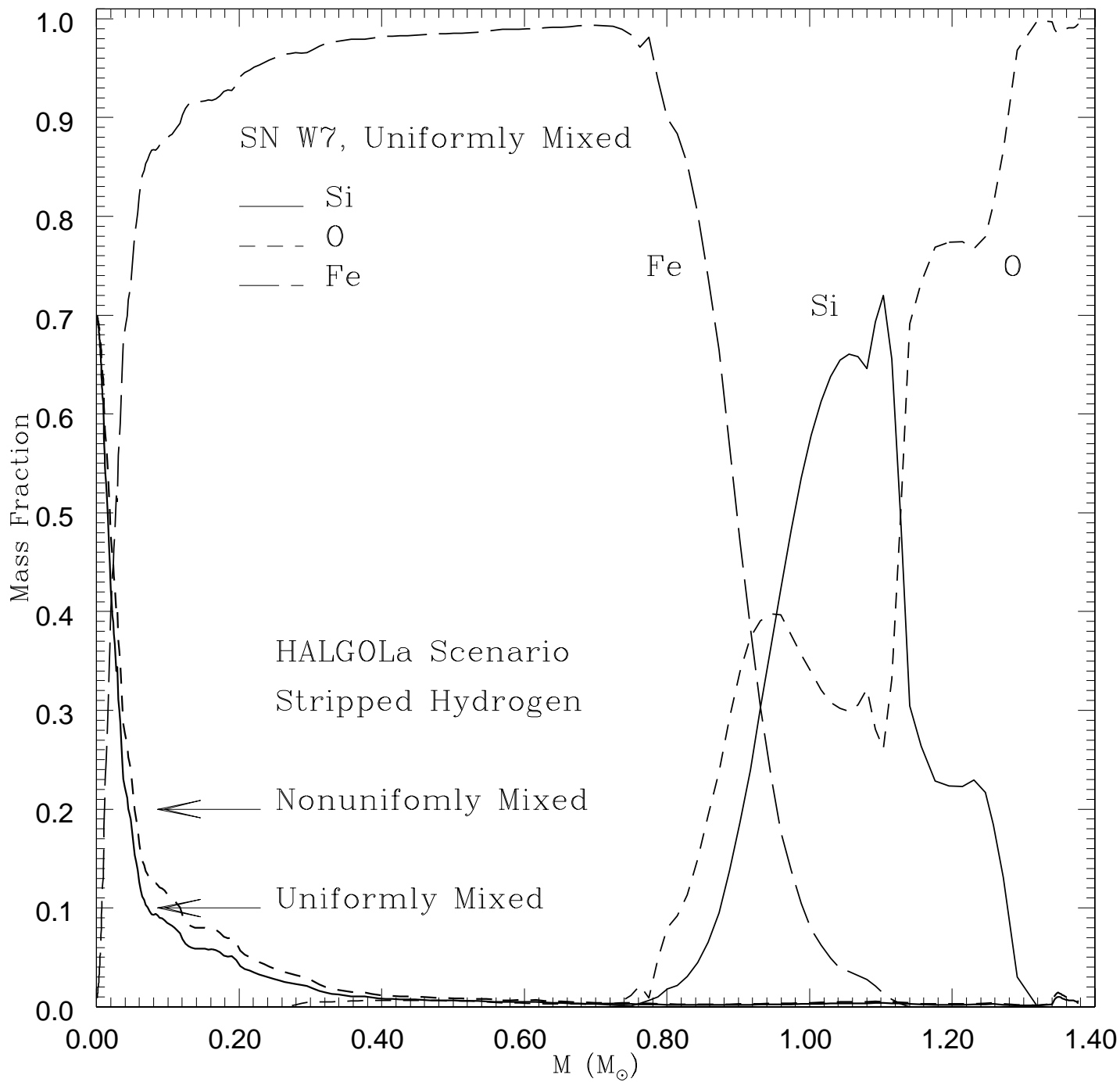
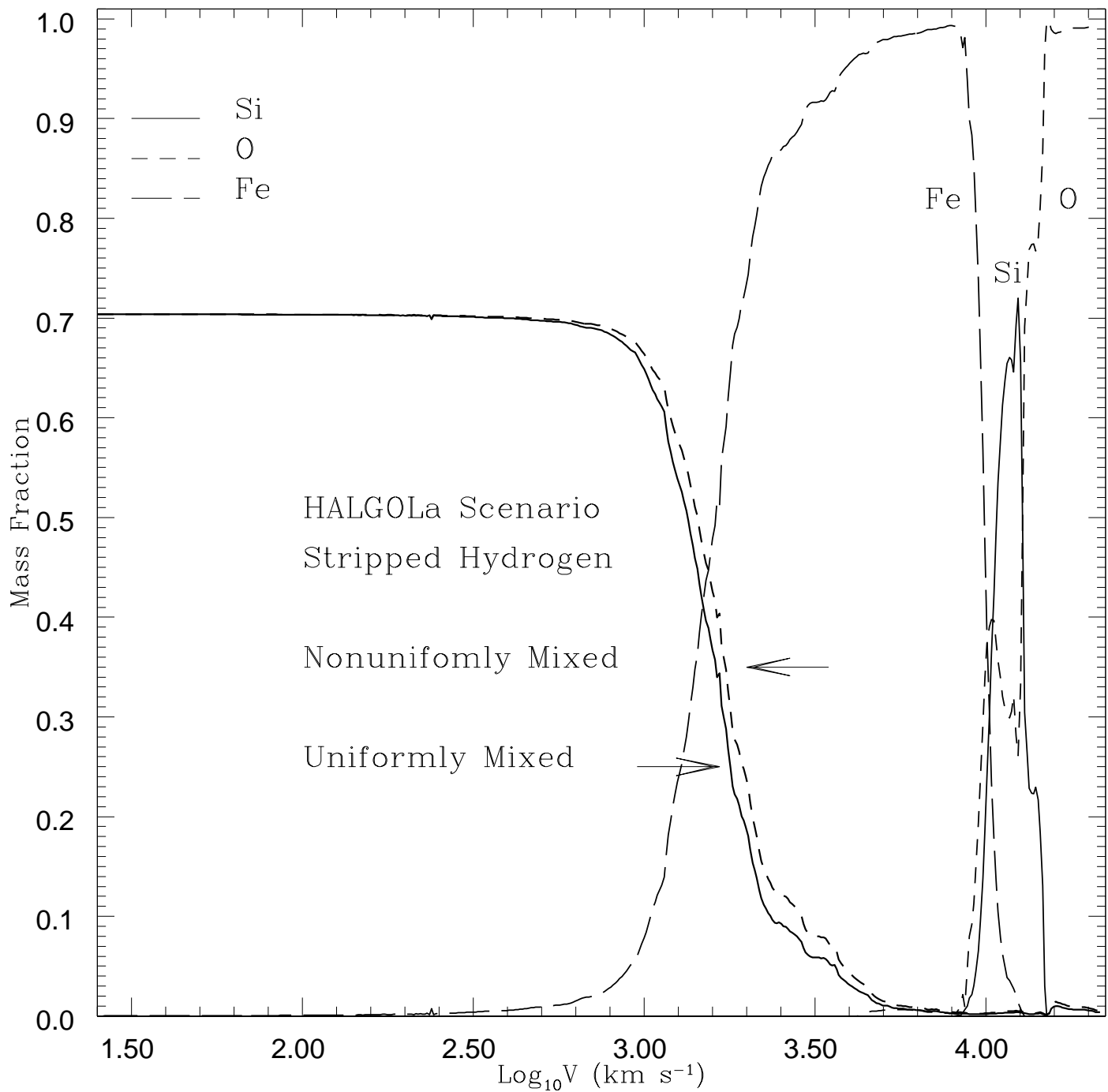
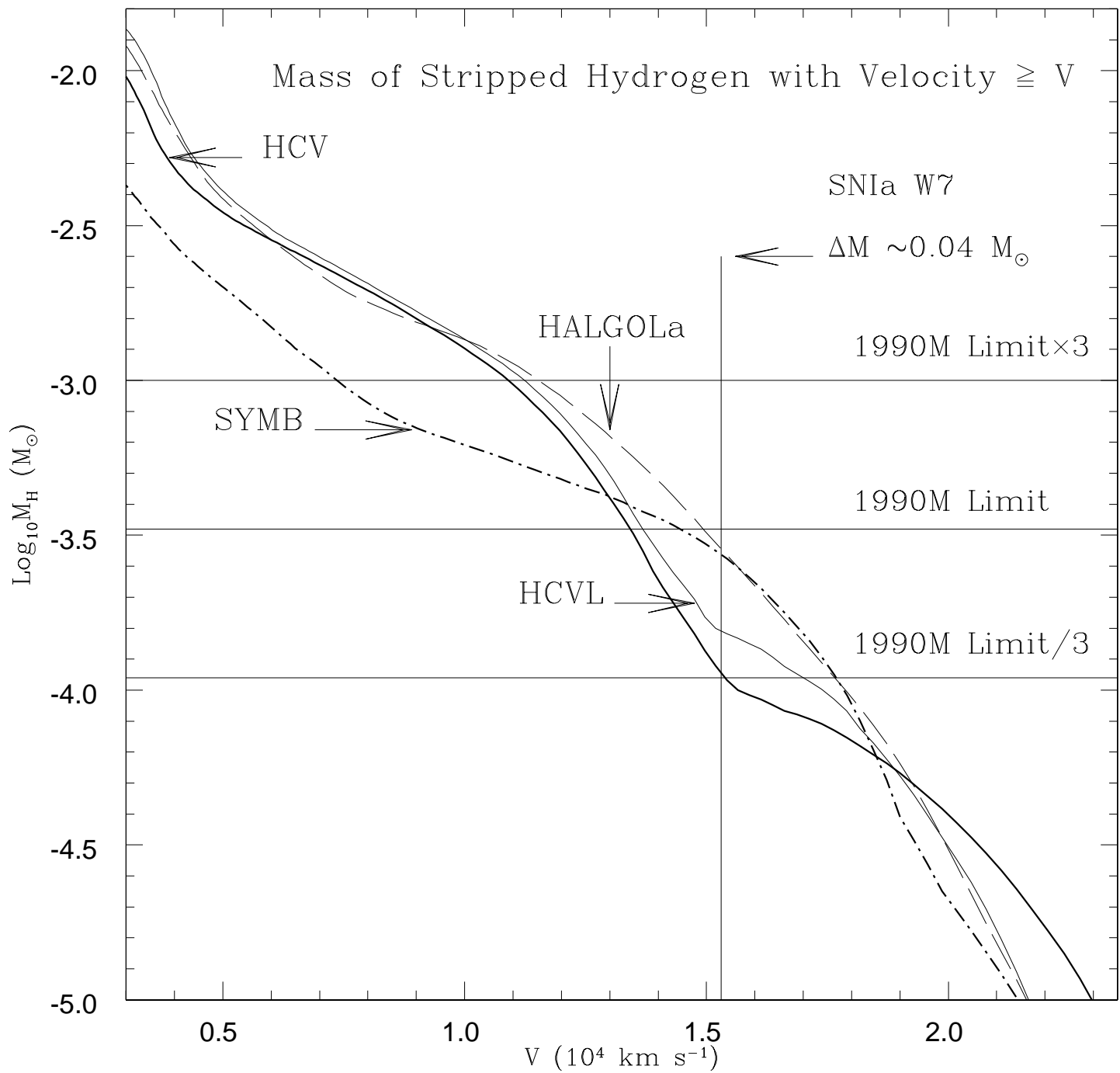
\end{document}